\documentclass[longauth]{aa}

\usepackage{graphicx}
\usepackage{txfonts}
\usepackage[
hidelinks,
colorlinks,
linkcolor=blue,
citecolor=blue]{hyperref}

\usepackage{graphicx}	
\usepackage{amsmath}	

\usepackage{caption} 
\usepackage[para,online,flushleft]{threeparttable}
\usepackage{rotating} 
\usepackage{booktabs}
\usepackage{url} 
\usepackage[table]{xcolor} 
\usepackage{bm}
\usepackage[flushleft]{threeparttable}

\usepackage{longtable}
\usepackage{pdflscape}
\usepackage{gensymb}

\usepackage[export]{adjustbox}

\begin{document}

\title{The ALMA survey to Resolve exoKuiper belt Substructures (ARKS)}
\subtitle{VIII: A dust arc and non-Keplerian gas kinematics in HD~121617}

\author{S. Marino\textsuperscript{1}\fnmsep\thanks{E-mail: s.marino-estay@exeter.ac.uk} \and V.~Gupta\textsuperscript{1} \and P.~Weber\textsuperscript{2,3,4} \and T.~D.~Pearce\textsuperscript{5} \and A.~Brennan\textsuperscript{6} \and S.~P\'erez\textsuperscript{2,3,4} \and S.~Mac~Manamon\textsuperscript{6} \and L.~Matr\`a\textsuperscript{6} \and J.~Milli\textsuperscript{7} \and M. Booth\textsuperscript{8} \and C.~del~Burgo\textsuperscript{9,10} \and G.~Cataldi\textsuperscript{11,12} \and E.~Chiang\textsuperscript{13} \and Y.~Han\textsuperscript{14} \and Th.~Henning\textsuperscript{15} \and A.~M.~Hughes\textsuperscript{16} \and M.~R.~Jankovic\textsuperscript{17} \and \'A.~K\'osp\'al\textsuperscript{18,19,15} \and J.~B.~Lovell\textsuperscript{20} \and P.~Luppe\textsuperscript{6} \and E.~Mansell\textsuperscript{16} \and M.~A.~MacGregor\textsuperscript{21} \and A.~Mo\'or\textsuperscript{18} \and J.~Olofsson\textsuperscript{22} \and A.~A.~Sefilian\textsuperscript{23} \and D.~J.~Wilner\textsuperscript{20} \and M.~C.~Wyatt\textsuperscript{24} \and B.~Zawadzki\textsuperscript{16}} 

\institute{
Department of Physics and Astronomy, University of Exeter, Stocker Road, Exeter EX4 4QL, UK \and
Departamento de Física, Universidad de Santiago de Chile, Av. V\'ictor Jara 3493, Santiago, Chile \and
Millennium Nucleus on Young Exoplanets and their Moons (YEMS), Chile \and
Center for Interdisciplinary Research in Astrophysics Space Exploration (CIRAS), Universidad de Santiago, Chile \and
Department of Physics, University of Warwick, Gibbet Hill Road, Coventry CV4 7AL, UK \and
School of Physics, Trinity College Dublin, the University of Dublin, College Green, Dublin 2, Ireland \and
Univ. Grenoble Alpes, CNRS, IPAG, F-38000 Grenoble, France \and
UK Astronomy Technology Centre, Royal Observatory Edinburgh, Blackford Hill, Edinburgh EH9 3HJ, UK \and
Instituto de Astrof\'isica de Canarias, Vía L\'actea S/N, La Laguna, E-38200, Tenerife, Spain \and
Departamento de Astrof\'isica, Universidad de La Laguna, La Laguna, E-38200, Tenerife, Spain \and
National Astronomical Observatory of Japan, Osawa 2-21-1, Mitaka, Tokyo 181-8588, Japan \and
Department of Astronomy, Graduate School of Science, The University of Tokyo, Tokyo 113-0033, Japan \and
Department of Astronomy, University of California, Berkeley, Berkeley, CA 94720-3411, USA \and
Division of Geological and Planetary Sciences, California Institute of Technology, 1200 E. California Blvd., Pasadena, CA 91125, USA \and
Max-Planck-Insitut f\"ur Astronomie, K\"onigstuhl 17, 69117 Heidelberg, Germany \and
Department of Astronomy, Van Vleck Observatory, Wesleyan University, 96 Foss Hill Dr., Middletown, CT, 06459, USA \and
Institute of Physics Belgrade, University of Belgrade, Pregrevica 118, 11080 Belgrade, Serbia \and
Konkoly Observatory, HUN-REN Research Centre for Astronomy and Earth Sciences, MTA Centre of Excellence, Konkoly-Thege Mikl\'os \'ut 15-17, 1121 Budapest, Hungary \and
Institute of Physics and Astronomy, ELTE E\"otv\"os Lor\'and University, P\'azm\'any P\'eter s\'et\'any 1/A, 1117 Budapest, Hungary \and
Center for Astrophysics | Harvard \& Smithsonian, 60 Garden St, Cambridge, MA 02138, USA \and
Department of Physics and Astronomy, Johns Hopkins University, 3400 N Charles Street, Baltimore, MD 21218, USA \and
European Southern Observatory, Karl-Schwarzschild-Strasse 2, 85748 Garching bei M\"unchen, Germany \and
Department of Astronomy and Steward Observatory, The University of Arizona, 933 North Cherry Ave, Tucson, AZ, 85721, USA \and
Institute of Astronomy, University of Cambridge, Madingley Road, Cambridge CB3 0HA, UK
}

\date{Received YYY; accepted XXX}

\abstract
   {ExoKuiper belts around young A-type stars often host CO gas, whose origin is still unclear. The ALMA survey to Resolve exoKuiper belt Substructures (ARKS) includes six of these gas-bearing belts, to characterise their dust and gas distributions and investigate the gas origin.
   }
   {As part of ARKS, we observed the gas-rich system HD~121617 with a 0\farcs12 (14~au) resolution and discovered an arc of enhanced dust density. In this paper,
we analyse in detail the dust and gas distributions and the gas kinematics of this system.
    }
   {We extracted radial and azimuthal profiles of the dust (in the millimetre and near-infrared) and gas emission ($^{12}$CO and $^{13}$CO) from reconstructed images. To constrain the morphology of the arc, we fitted an asymmetric model to the dust emission. To characterise the gas kinematics, we fitted a Keplerian model to the velocity map and extracted the gas azimuthal velocity profile by deprojecting the data.  
   }
   {We find that the dust arc is narrow (1--5~au wide at a radius of 75~au), azimuthally extended with a full width at half maximum of ${\sim}90\degree$, and asymmetric; the emission is more azimuthally compact in the direction of the system's rotation, and represents 13\% of the total dust mass (0.2~$M_{\oplus}$). From analysis of the scattered light and CO images, we conclude that the arc is much less pronounced or absent for small grains and gas. Finally, we find strong non-Keplerian azimuthal velocities at the inner and outer wings of the ring, as was expected due to strong pressure gradients.}
   {The dust arc resembles the asymmetries found in protoplanetary discs, often interpreted as the result of dust trapping in vortices. If the gas disc mass is high enough ($\gtrsim20~M_{\oplus}$, requiring a primordial gas origin), both the radial confinement of the ring and the azimuthal arc may result from dust grains responding to gas drag. Alternatively, it could result from planet-disc interactions via mean motion resonances. Further studies should test these hypotheses and may provide a dynamical gas mass estimate in this CO-rich exoKuiper belt. }

\keywords{Planetary systems; Submillimeter:planetary systems; Circumstellar matter; Surveys; Techniques:interferometric
               }

\maketitle

\section{Introduction}

ALMA observations of debris discs, in particular those analogous to the Kuiper belt (exoKuiper belts), have the potential to constrain the formation and architecture of planetary material at tens of astronomical units \citep{Miller2021, Najita2022, Pearce2022}. The recent REASONS survey \citep{Matra2025} revealed a great diversity in the morphology of exoKuiper belts using ALMA observations. However, the low resolution of these observations hampered a systematic study of substructures in these discs. The ALMA survey to Resolve exoKuiper belt Substructures \citep[ARKS,][]{overview_arks} is the first ALMA Large Program dedicated to studying substructures in the dust and gas of 24 exoKuiper belts. One of the key findings of ARKS is the presence of significant asymmetries in 10/24 belts \citep{asym_arks}.

Among the belts in ARKS, six belong to the particular class of gas-bearing debris discs that are frequently found around young A-type stars \citep[][]{Kospal2013, Lieman-Sifry2016, Moor2017, Cataldi2023}. The origin of gas (CO, atomic carbon, and other atomic species) is uncertain in these discs. On the one hand, it could be residual protoplanetary disc gas that has not yet dispersed and is dominated by molecular hydrogen \citep{Nakatani2021, Nakatani2023, Ooyama2025}. On the other hand, the CO gas could be released from the solids via thermal desorption, photodesorption, or collisions, and be dominated by volatiles such as CO, carbon, and oxygen \citep{Zuckerman2012, Dent2014, Marino2016,  Kral2019, Marino2020gas, Marino2022, Bonsor2023}.

Of the ten asymmetric belts in ARKS, HD~121617 stands out. This system is composed of a 16~Myr old A1V type star at 118~pc \citep{Pecaut2016, Houk1978, gaiadr3} that hosts the brightest gas disc in the sample \citep{gas_arks}, and displays an asymmetry in the form of a dust overdensity or arc \citep{asym_arks}. Its debris disc was first identified as an infrared excess detected from the mid-infrared with IRAS, AKARI, and WISE to the far-infrared with Herschel \citep{Mannings1998, Fujiwara2013, Perrot2023}. It was later resolved with ALMA at low resolution \citep{Moor2017} and more recently in scattered light as a narrow dusty debris disc \citep{Perrot2023}. The disc has a fractional luminosity of $5\times10^{-3}$ \citep{overview_arks}, which is among the highest in the sample of known gas-rich debris discs \citep{Moor2017, Moor2025}.

The scattered light data, and their re-analysis as part of ARKS, revealed a narrow belt of small dust grains and a potential small eccentricity of 0.03, with a pericentre towards the southwest (SW) ansa where the dust arc peaks \citep{Perrot2023, scat_arks}. The new ALMA observations revealed an azimuthally extended overdensity in millimetre dust \citep{asym_arks} that resembles the overdensities seen in protoplanetary discs, typically attributed to dust trapped in vortices \citep[e.g.][]{Casassus2013, vandermarel2013, Marino2015, Baruteau2016, Regaly2017}. The CO gas emission, on the other hand, does not show any strong asymmetry \citep{gas_arks}. HD~121617's low inclination and high CO flux make it the ideal ARKS target to study dust and gas asymmetries, as well as gas kinematics in exoKuiper belts.

\begin{figure*}
    \centering
    \includegraphics[width=1.0\textwidth]{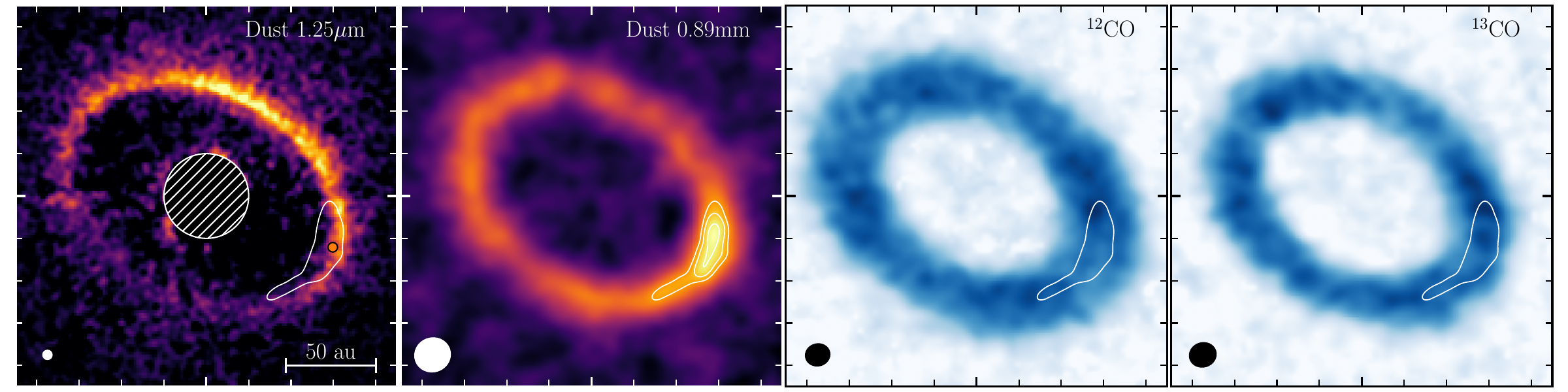}
    \caption{Dust and gas images of HD~121617. Left panel: Scattered light VLT/SPHERE $Q_\phi$ image at 1.25~$\mu$m ($J$ band) observed in polarised light with IRDIS \citep{scat_arks}, smoothed with a Gaussian with a standard deviation of 1 pixel (12~mas). The central hatched region masks an area dominated by strong artefacts. The small orange circle in the SW ansa marks the pericentre location as constrained by \cite{scat_arks}. Middle left panel: ALMA dust continuum image at 0.89~mm \citep{overview_arks}. Middle right panel: $^{12}$CO J=3-2 moment 0 image \citep[][]{gas_arks}. Right panel: $^{13}$CO J=3-2 moment 0 image \citep{gas_arks}. The white contours in the middle left image represent 75, 85, and 95\% of the peak intensity. The 75\% contour is also shown in the other images for reference. The ALMA continuum and CO images were obtained with CLEAN using robust parameters of 1.0 and 0.5, respectively. The beam and PSF sizes are shown as ellipses in the bottom left of each panel. The minor ticks in all panels are spaced by 0\farcs2.}
    \label{fig:images}
\end{figure*}

Motivated by the dust overdensity, this paper closely examines the dust asymmetry and gas kinematics to assess the nature of this feature. In Sect.~\ref{sec:dist}, we examine the dust and gas distributions and fit a parametric model to the millimetre dust distribution to constrain the morphology of the arc. In Sect.~\ref{sec:kinematics}, we analyse the $^{12}$CO J=3-2 (hereafter $^{12}$CO, unless otherwise stated) gas kinematics and find strong deviations from Keplerian rotation using the ARKS new high spectral (velocity) resolution. Sect.~\ref{sec:discussion} discusses the origin of the arc, the presence of non-Keplerian kinematics in exoKuiper belts, and the gas origin in this system. Finally, Sect.~\ref{sec:conclusions} summarises our findings and conclusions.

\section{Dust and gas distribution}
\label{sec:dist}

To constrain the distribution of large millimetre-sized grains and CO gas, in this section we use the new ALMA images obtained as part of ARKS (2022.1.00338.L, PI: S. Marino, co-PIs: A.~M.~Hughes \& L.~Matr\`a). HD~121617 was observed in band 7 (0.89~mm) between October 2022 and May 2023 using two antenna configurations to sample small and large scale structures, which yielded a resolution of $0\farcs12$ or 14~au and a maximum recoverable scale of ${\sim}5$\arcsec\ or 600~au. The observations included four spectral windows targeting the dust continuum emission that is dominated by millimetre-sized grains and the $^{12}$CO and $^{13}$CO gas J=3-2 emission. The $^{12}$CO and $^{13}$CO observations had a spectral resolution of 26 and 850~m/s, respectively. This means that only the $^{12}$CO data is suitable for a detailed analysis of its kinematics. Imaging of the continuum and CO data was performed using the Clean algorithm \citep[for more details see][]{overview_arks, gas_arks}.  

To constrain the distribution of small grains, we used VLT/SPHERE polarised data that traces the scattered light by small micron-sized grains. We used a re-reduced $Q_\phi$ image at 1.25~$\mu$m ($J$ band), obtained using IRDIS on 18 April 2018 and achieving a resolution of $0\farcs04$ or 5~au. The data were originally presented in \cite{Perrot2023} and re-reduced by \cite{scat_arks} as part of ARKS. 

The SPHERE and ALMA images are shown in Figure~\ref{fig:images}. The peak signal-to-noise (S/N) of the SPHERE image, ALMA dust, $^{12}$CO, and $^{13}$CO images is 10, 24, 21, and 19, respectively. The asymmetry is only strong in the dust emission at 0.89~mm, which is highlighted with white contours in all images. Note that the increased emission towards the NW in the scattered light image at 1.25~$\mu$m is due to the polarised scattering phase function, which favours forward scattering, i.e. scattering from the near-side of the belt \citep{Min2012}.

\subsection{Radial and azimuthal profiles}
\label{sec:profiles}

Here, we focus on characterising the distribution of the micron-sized grains, millimetre-sized grains, and CO gas uniformly. For this, we extracted radial and azimuthal intensity profiles from the SPHERE scattered light image, ARKS dust continuum image, and ARKS $^{12}$CO and $^{13}$CO J=3-2 moment 0 images.

\begin{figure*}
    \centering
    \includegraphics[width=0.9\textwidth]{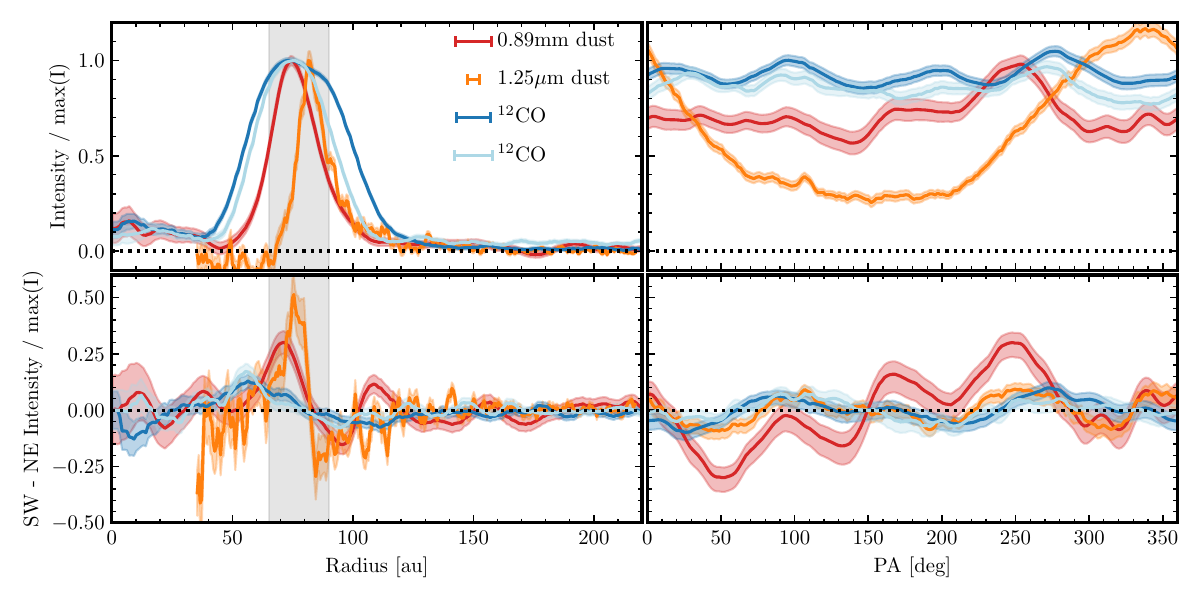}
    \caption{Radial and azimuthal profiles extracted from the SPHERE and ALMA dust and gas images (produced with a robust parameter of 0.5). Top left panel: De-projected and azimuthally averaged intensity radial profiles extracted from wedges towards the NE and SW ansae with a width of 120\degree\, for ALMA and 60\degree\, for SPHERE. The bars on the top right indicate the resolution of each profile. Bottom left: Residual radial profiles after subtracting the average intensity of the NE from the SW ansa. Top right: Intensity azimuthal profiles obtained by averaging the emission between 65 and 90~au and over 40\degree\, rolling windows. Bottom right: residual azimuthal profiles after subtracting the mirrored profile relative to the NW minor axis of the disc (PA of 330\degree). The shaded coloured region in all panels represents the $1\sigma$ uncertainty taking into account the image noise and number of independent points being averaged (i.e. points spaced by a beam). The vertical shaded grey region in the left panels shows the radial region that was averaged when extracting the azimuthal profiles. The dust and gas intensities are normalised by their radial profile peaks.}
    \label{fig:profiles}
\end{figure*}

The top left panel of Figure~\ref{fig:profiles} shows the radial profiles extracted directly from the CLEAN images using the same approach as presented in \cite{rad_arks}\footnote{For the ALMA dust continuum and CO images we use two opposing 120\degree\ wide wedges aligned with the disc major axis, whereas for the SPHERE image we use narrower wedges of 60\degree\ since the emission is affected by the phase function.} and assuming the images to be well centred on the stars\footnote{The stellar location is well known in the SPHERE observations and corresponds to the image centre \citep{scat_arks}. For ALMA, the star is not detected. However, from Gaia DR3's astrometric solution and ALMA astrometric precision, we know the star should be at the centre with an uncertainty of ${\lesssim}10$~mas \citep{overview_arks}.}. As is discussed in \cite{scat_arks}, the small dust distribution traced at 1.25~$\mu$m peaks ${\sim}7$~au further out than the millimetre emission, which could be due to the combined effect of radiation pressure and gas-drag \citep[as explored in][]{hd131835_arks}. The $^{12}$CO emission profile, on the other hand, peaks at a radius of 73~au that is similar to the peak radius of large dust grains (74~au). However, the CO emission is more radially extended with a full width at half maximum (FWHM) of 48~au \citep{gas_arks} compared to the 12~au for the small and 22~au for the large dust components (as measured from the radial profiles in Fig.~\ref{fig:profiles}). The $^{13}$CO emission is less extended with a FWHM of 37~au compared to the $^{12}$CO, although still more extended than both the millimetre- and micron-sized dust profiles. The wider radial span of $^{12}$CO compared to $^{13}$CO and the dust can be simply explained by its high optical depth as demonstrated in \cite[][]{line_arks}, with the surface density profile having a width that is consistent with that of the dust.

In addition, we investigate whether the emission profiles along the NE and SW ansae display any shift due to the belt being eccentric, as found in the scattered light emission \citep{Perrot2023, scat_arks}. The bottom left panel of Figure~\ref{fig:profiles} shows the difference between radial profiles extracted from wedges towards the SW and NE ansae, hereafter called residuals. The residuals from the scattered light (orange) display an `S' profile, indicating that the SW emission is slightly closer to the star relative to the NE emission. The fact that the positive and negative have roughly the same amplitude implies that there is no strong excess of emission on the SW side of the scattered light, in contrast to the millimetre emission, but an offset in position. Because the SW is slightly closer to the star than the NE due to the 0.03 eccentricity, the SW side would be expected to be ${\sim}10$\% brighter in scattered light, which may explain why the positive residual has a larger amplitude than its negative counterpart in the scattered light profile. 

The $^{12}$CO and $^{13}$CO residuals also display a significant signal with an `S' profile as the scattered light profile, indicating a potential eccentricity in the CO emission. The $^{12}$CO emission peaks at 80~au on the NE side versus 75~au on the SW side. This difference is consistent with that expected from an eccentricity of ${\sim}0.03$. The residual is still significant if we assume a different stellar location; for example, the centre derived from the millimetre emission (Sect.~\ref{sec:model}) or the $^{12}$CO kinematics (Sect.~\ref{sec:kinematics}). However, we note that the stellar location uncertainty of ${\lesssim}10$~mas \citep{overview_arks} is of the same order as the expected belt offset from a 0.03 eccentricity (20~mas) towards the NE direction. Additionally, an apparent offset in the minor axis direction would be expected if the $^{12}$CO and $^{13}$CO emission is optically thick and tracing a vertically elevated $\tau\sim2/3$ layer \citep{line_arks}, as is seen for protoplanetary discs. In that case, the offset would differ between $^{12}$CO and $^{13}$CO due to the different heights of this layer, but this difference is too small to be disentangled with the current uncertainties. Therefore, we cannot confirm this eccentricity in the CO emission.

The millimetre dust residual displays a peak that corresponds to the arc location. Beyond the positive residual at 75~au, we find a $3\sigma$ negative residual at 95~au. This suggests that the millimetre dust distribution could also be eccentric, with the NE side being more extended. However, the significance of this feature drops to $2.5\sigma$ if we assume a stellar location at the centre derived from the millimetre emission fit (Sect.~\ref{sec:model}). The SW profile peaks at 74.2 au and NE at 75.2, which is a difference much smaller than expected for the 0.03 eccentricity seen in scattered light. However, the fact that the millimetre emission is highly asymmetric makes it challenging to assess whether the disc is eccentric. Therefore, we can neither confirm nor rule out that the millimetre emission is eccentric.

The top right panel of Figure~\ref{fig:profiles} shows the azimuthal intensity profiles with a PA of 0\degree\ aligned with north increasing anti-clockwise on the plane of the sky. The profiles are extracted by averaging the emission between 65 and 90~au (roughly the FWHM of the millimetre emission) using a rolling azimuthal window of 40\degree\ projected in the sky plane, which boosts the S/N and smooths the recovered profile. The asymmetry in the millimetre dust emission (solid red line) peaks at a position angle (PA) of ${\sim}250\degree$, but it spans a wide range of PA. The peak excess is about ${\sim}40\%$ of the average intensity beyond the arc. However, we note that both the beam and azimuthal averaging could make this excess appear lower than its true value. For example, our modelling in Sect.~\ref{sec:model} shows that the overdensity could have a peak density that is twice the background density or even larger. Finally, the excess seems to have a sharper edge towards larger PAs (anti-clockwise from the peak) than to the opposite side.

The scattered light emission (solid orange line) has a strong azimuthal dependence due to the polarised scattering phase function that favours forward-scattering. To account for this effect, we self-subtracted the emission mirroring the azimuthal profile with respect to the semi-minor axis at the near side of the ring \citep[at a PA of 330\degree as derived from scattered light,][and consistent with the ALMA continuum and CO gas kinematics modelling in Sect.~\ref{sec:model} and Sect.~\ref{sec:kinematics}]{scat_arks}. The self-subtracted profile (orange line in the bottom right panel of Figure~\ref{fig:profiles}) shows a $2\sigma$ excess centred at ${\sim}250\degree$, and that is ${\sim}20\%$ of the emission at the corresponding PA. However, this feature only appears when averaging the emission between 65 and 90~au. If we extend this range to 110~au, this asymmetry disappears. This is because the scattered light emission is slightly eccentric as described above, with the SW side being closer in than the NE side \citep{scat_arks}. Overall, these findings suggest that the arc, if any, is less prominent in the distribution of small grains.

The $^{12}$CO and $^{13}$CO emission displays an azimuthal modulation with four maxima 45\degree\ away from the major and minor axes. This is common in moment 0 maps for optically thick line emission, as the Keplerian shear along the line of sight and within a beam makes the line wider and the integrated line flux higher in those locations \citep{line_arks}. Apart from this modulation, we do not find any significant feature (e.g. the residual azimuthal profile is consistent with zero). We also analyse the azimuthal profile of the moment 8 maps (i.e. peak intensity image) and find no significant excess at any PA. Therefore, we conclude that there is no significant asymmetry in the $^{12}$CO and $^{13}$CO emission (apart from a small eccentricity). However, since both the $^{12}$CO and $^{13}$CO emissions are optically thick, we cannot rule out the presence of an asymmetry in the gas density distribution similar to that seen in the millimetre-sized dust.

\subsection{Arc parametric fit to continuum emission}
\label{sec:model}

\begin{figure*}
    \centering
    \includegraphics[width=0.8\linewidth]{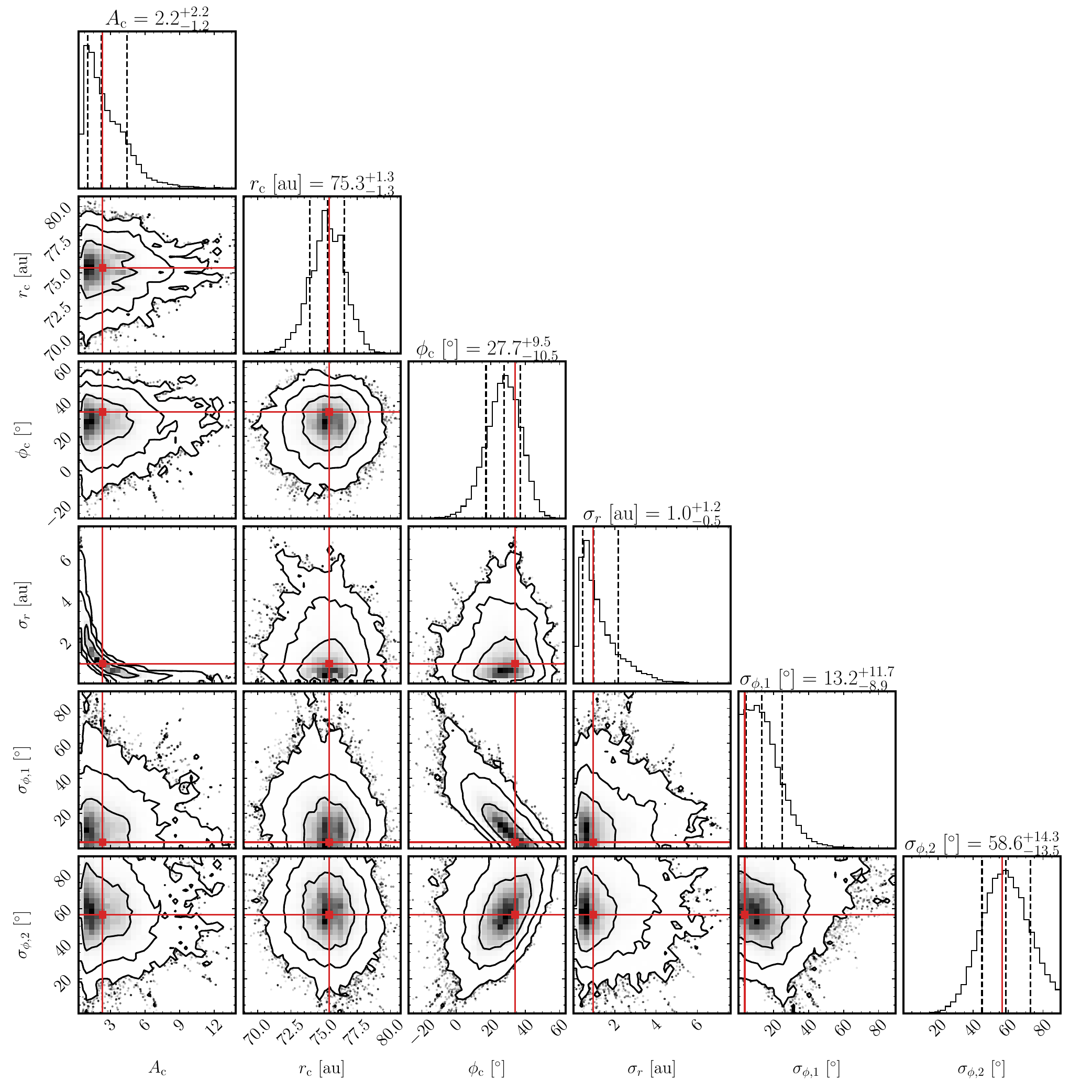}
    \caption{MCMC posterior distribution of the arc parameters used to fit the dust continuum observations: the peak amplitude of the arc $A_{\rm c}$; the central radius of the arc $r_{\rm c}$; the azimuthal location of the arc peak $\phi_{\rm c}$; the arc radial standard deviation $\sigma_r$; and the azimuthal standard deviations for the leading and trailing side of the arc $\sigma_{\phi,1}$ and $\sigma_{\phi,2}$, respectively. The contour levels in the 2D marginalised distributions correspond to the 68, 95 and 99.7\% confidence levels. The dashed vertical lines in the marginalised distributions display the 16th, 50th and 84th percentiles. The red lines represent the best-fit value (lowest $\chi^{2}$). The posterior distribution of the other parameters is presented in Figure~\ref{fig:full_mcmc}}.
    \label{fig:mcmc}
\end{figure*}

To constrain the morphology of the arc in the distribution of millimetre-sized grains, we fitted a parametric model to the continuum visibilities following the same approach as in \cite{overview_arks} using \href{https://github.com/SebaMarino/disc2radmc}{disc2radmc} \citep[a python wrapper to use RADMC3D\footnote{https://www.ita.uni-heidelberg.de/~dullemond/software/radmc-3d/},][]{Marino2022} and an MCMC to sample the parameter space. Simply put, we used a parametric model to define the disc density distribution, produced a synthetic image using radiative transfer calculations, Fourier transformed the image and calculated the model likelihood using Galario \citep{galario}, and sampled the posterior distribution using the Python package \textsc{emcee} \citep{emcee} to fit the free parameters using the Affine Invariant MCMC Ensemble sampler \citep{Goodman2010}.

We modelled the dust surface density as the sum of two components: an axisymmetric component, $\Sigma_a(r)$, described as a double power law as a function of the radial cylindrical coordinate, $r$, which was found to provide a good fit to the azimuthally averaged radial profile \citep{rad_arks}; and an asymmetric component $\Sigma_{\rm c}(r,\phi)$ described as a 2D Gaussian as a function of $r$ and the azimuthal angle, $\phi$, measured anti-clockwise in the plane of the disc from the SW ansa. This parametrisation offers adequate flexibility to constrain the radial and azimuthal widths of the arc. The surface density is thus defined as
\begin{eqnarray}
\Sigma(r, \phi) &=&  \Sigma_a(r)+\Sigma_{\rm c}(r,\phi),\\
\Sigma_a(r) &=& 2^{1/\gamma} \Sigma_0 \left[\left(\frac{r}{r_0}\right)^{-\alpha_{\rm in}\gamma} + \left(\frac{r}{r_0}\right)^{-\alpha_{\rm out}\gamma} \right]^{-1/\gamma}, \\
\Sigma_{\rm c}(r, \phi)&=&  A_{c} \Sigma_0 \exp\left[- \frac{(r-r_{\rm c})^2}{2\sigma_r^2} - \frac{(\phi-\phi_{\rm c})^2}{2\sigma_\phi^2}\right],
\\
\sigma_\phi&=&
\begin{cases}
\sigma_{\phi1}, &\text{if }  \phi>\phi_c \\
\sigma_{\phi2}, &\text{if }  \phi<\phi_c
\end{cases}
\end{eqnarray}
where $\alpha_{\rm in}$ and $\alpha_{\rm out}$ control how steep the inner and outer edges of the belt are, $\gamma$ controls how smooth or sharp the surface density radial peak is, and $r_0$ determines the peak radius of the axisymmetric component. For the asymmetric component, $A_{\rm c}$ controls the peak density of the arc relative to the axisymmetric component, $r_{\rm c}$ and $\sigma_r$ control the radial location and width of the arc, whereas $\phi_{\rm c}$ and $\sigma_\phi$ control its azimuthal location and width. 

Furthermore, we split $\sigma_\phi$ into two parameters to allow the arc to be azimuthally asymmetric as the azimuthal profile suggests. We thus use $\sigma_{\phi1}$ for $\phi>\phi_{\rm c}$  and $\sigma_{\phi2}$ for $\phi<\phi_{\rm c}$. From the scattered light image, we can infer that the NW side is the near side, and from the CO line-of-sight velocities, the SW has a projected velocity towards us. Therefore, the disc is rotating counterclockwise. This means that $\sigma_{\phi1}$ determines the azimuthal width in the leading side of the arc, while $\sigma_{\phi2}$ does for the trailing side. Finally, we also leave as free parameters the disc inclination and position angle (longitude of ascending node), and RA and Dec offsets. 

As in \cite{overview_arks}, we modelled the dust as a size distribution from 1~$\mu$m to 1~cm, with a power law exponent of -3.5. 
We assumed a dust composition made of 70\% astrosilicates, 15\% crystalline water ice, and 15\% amorphous carbon by mass \citep[mixed using the Bruggeman's formula,][]{Draine2003, LiGreenberg1998}. We assumed grains to be compact and spherical and used Mie theory to calculate their opacities. This results in an opacity of 1.9~cm$^{2}$~g$^{-1}$  at 0.89~mm (the only wavelength being fitted here). The minimum grain size may be smaller, as inferred from the SED, or larger if equal to the blow-out size \citep[${\sim}3\ \mu$m,][]{Perrot2023}. Nevertheless, it does not affect the derived dust masses, which are dominated by the maximum grain size given the assumed size distribution exponent. The maximum size of solids in the belt is likely much larger than 1~cm (especially if planetesimals are present), meaning that the derived dust mass only accounts for grains smaller than 1~cm, which dominate the millimetre emission \citep{Terrill2023}.

The best-fit parameters for the arc and their posterior distribution are presented in Figure~\ref{fig:mcmc} (the posterior distribution of the remaining parameters is presented in Figure~\ref{fig:full_mcmc}). For the axisymmetric component, we find it is centred at $r_0=70\pm1$~au, with an inner slope $\alpha_1>10$ ($3\sigma$) and outer slope $\alpha_2=-6.5\pm0.3$, consistent with the values derived in \cite{rad_arks}. The peak of the surface density is likely sharp, but it is only marginally constrained by $\gamma$. Combined, these parameters set the surface density peak of the axisymmetric component at $70\pm1$~au. 

The total dust mass is constrained to be $0.233\pm0.004\ M_{\oplus}$, but we note that the small uncertainties do not include the uncertainties in the assumed dust opacity. The derived value agrees well with the dust mass of $0.21\pm0.02\ M_{\oplus}$ derived by \cite{Perrot2023} fitting the spectral energy distribution. However, these dust masses are not directly comparable since there are subtle differences between the size distribution parameters (e.g. slope and maximum grain size) and composition between both modelling efforts. Depending on the assumed dust opacity, the derived dust mass could vary by a factor of a few \citep[e.g.][]{Kim2019}. Despite these large systematic uncertainties, our results and conclusions do not depend on the exact value of the dust mass.

For the asymmetric component, we find that the arc peak, $r_{\rm c}$, is centred at $75\pm1$~au (i.e. slightly further out than the axisymmetric component) at an azimuth, $\phi_{\rm c}$, of $28^{+10}_{-11}$\degree, which translates to a PA of ${\sim}$270\degree, given the disc inclination and longitude of ascending node. The radial width of the arc, $\sigma_r$, is marginally resolved with a standard deviation of $1.0^{+1.2}_{-0.5}$~au (a FWHM of $2.4^{+2.8}_{-1.2}$~au) and a 3$\sigma$ upper limit of 5.4~au. Azimuthally, we find strong evidence that the arc's trailing side is wider, with a standard deviation of $59\pm14$\degree\ ($\sigma_{\phi,2}$) on the trailing side versus $13^{+12}_{-9}$\degree\ on the leading side ($\sigma_{\phi,1}$). Using the posterior distribution, we estimate the azimuthal FWHM ($2.355(\sigma_{\phi,1}+\sigma_{\phi,2})/2$) to be $85^{+18}_{-21}$\degree, confirming the wide azimuthal extension of this dust overdensity. Finally, the arc peak amplitude, $A_{\rm c}$, is not well constrained as its value is degenerate with the radial width, which is not well constrained. However, from the posterior distribution, we calculated and found that the overall millimetre dust mass contribution of the arc is $13\pm1\%$ of the total dust mass of 0.23~$M_{\oplus}$.

\begin{figure}
    \centering
    \includegraphics[width=1.0\columnwidth]{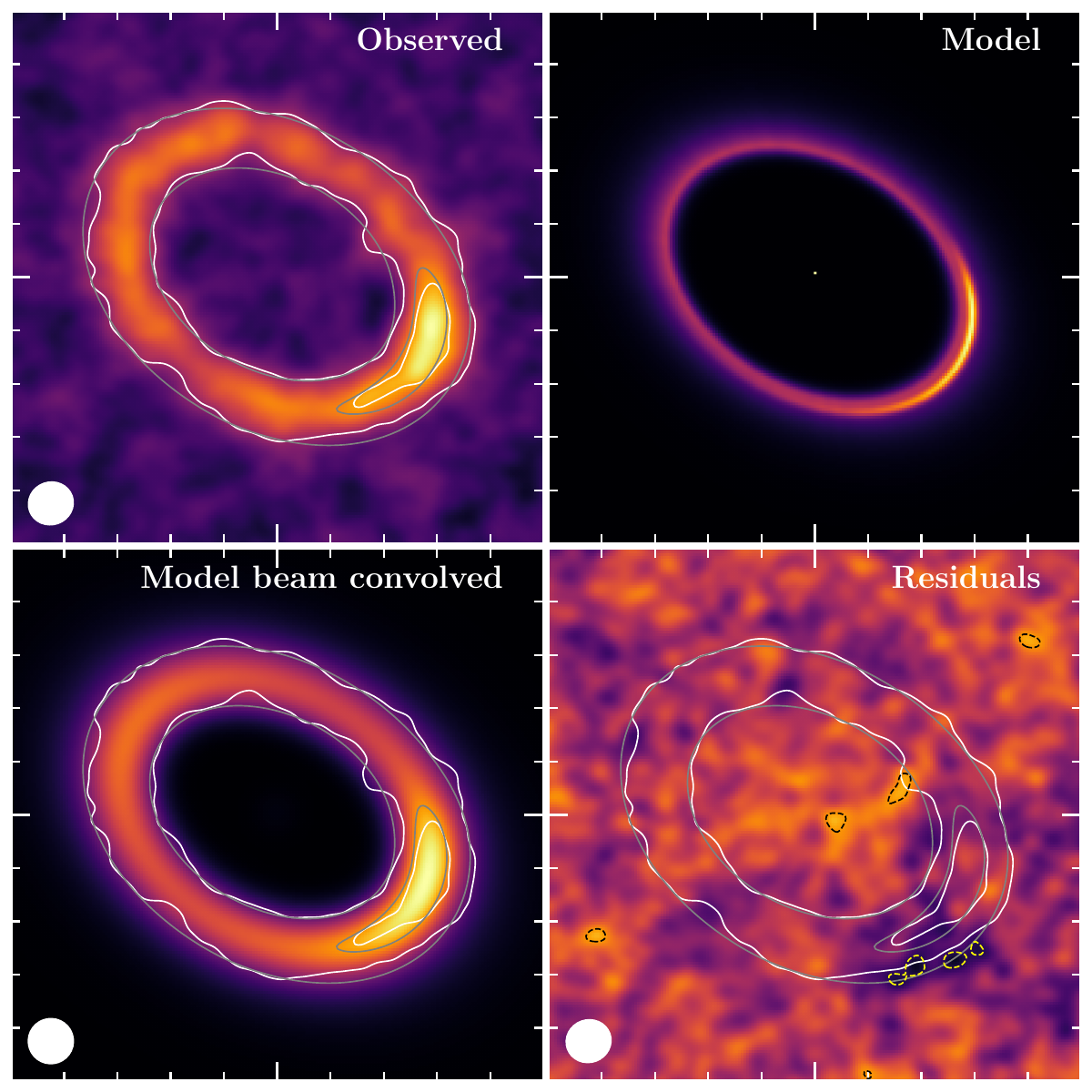}
    \caption{Comparison between observations and best-fit model. Top left: ALMA dust continuum image. Top right: Synthetic image of the best-fit model. Bottom left: Best-fit model image that has been beam-convolved. Bottom right: Dirty image of the residuals after subtracting the best-fit model from the observed visibilities. The dashed black and yellow contours represent positive and negative $3\sigma$ values, respectively. The white (grey) contours represent an intensity that is 30 and 75\% the intensity peak of the observed (convolved model) image. The beam sizes are displayed in the bottom left corners. All images correspond to a CLEAN robust parameter of 1.0. The small white ticks are spaced by 0.2".  }
    \label{fig:model}
\end{figure}

Figure~\ref{fig:model} shows the ALMA continuum image, the model image, the beam-convolved model image, and the residuals. Overall, we find a good match between the observed morphology and the model. The only significant discrepancy is found at the SW side just exterior to the trailing side of the arc, where the residual map shows a few $3\sigma$ negatives (dashed yellow contours). The circular model is more extended than the data in this area of the disc. This may be related to the eccentricity found in scattered light with a pericentre towards the SW and the slight negatives in the residuals near 100~au found in the SW-NE subtracted radial profiles. These findings suggest that the millimetre dust distribution may be eccentric, too. Alternatively, the arc radial distribution may be more complex than our model; for example, with the trailing side being closer to the star.

\section{CO gas kinematics}
\label{sec:kinematics}

In addition to constraining the CO gas distribution in Sect.~\ref{sec:profiles}, here we investigate the kinematics of the CO gas traced by the $^{12}$CO ARKS observations with a resolution of 26 m/s (13~m/s channels). We use the clean data cube corresponding to a robust value of 0.5 \citep{gas_arks}.  

\subsection{2D velocity map}

\begin{figure}
    \centering
    \includegraphics[trim=0.0cm 0.0cm 0.2cm 0.2cm,
    clip=true,width=1.0\linewidth]{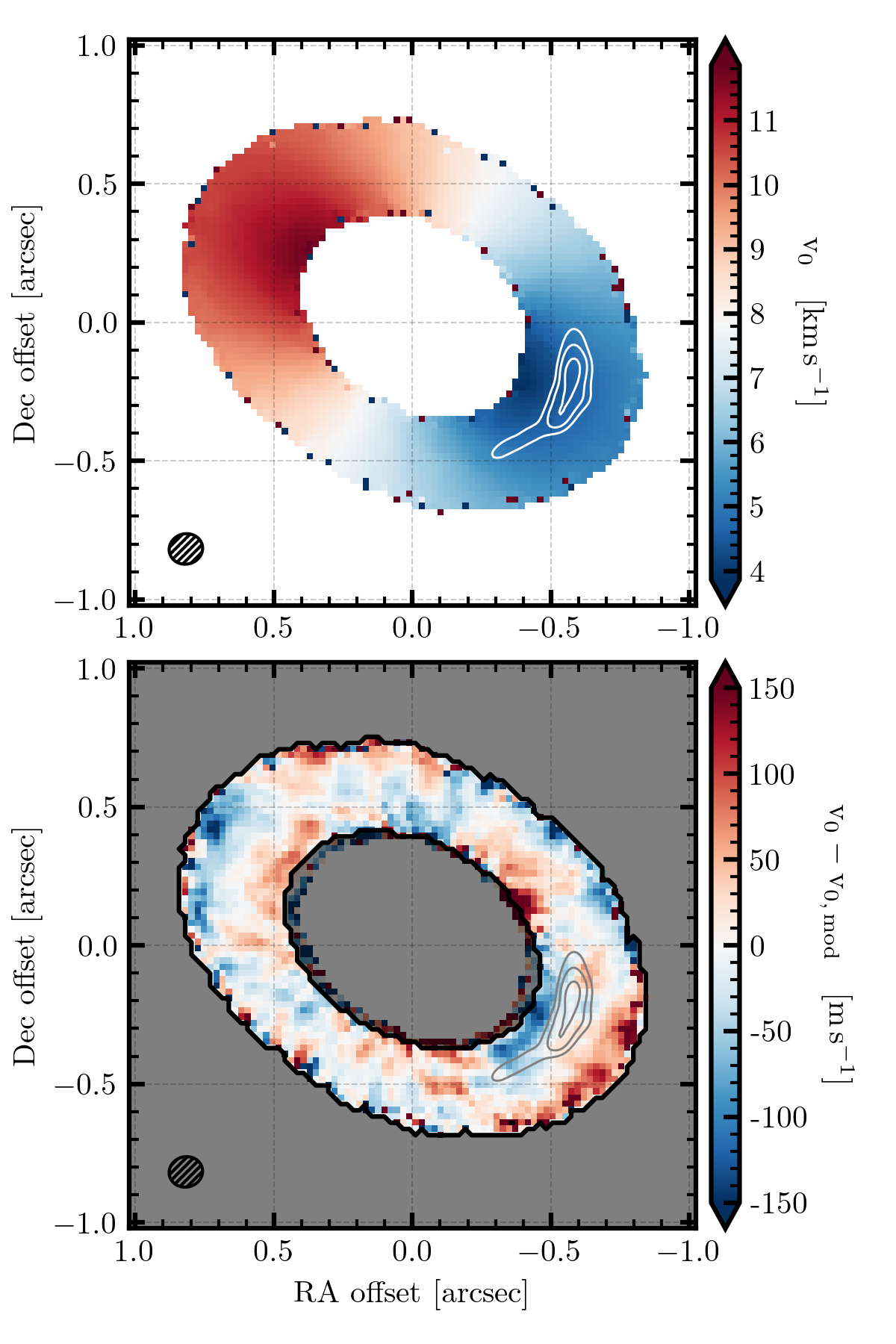}
    \caption{Top panel: Line of sight $^{12}$CO velocity map. Bottom panel: Residual velocity map after subtracting a Keplerian model. The white-shaded region in the top panel represents pixels that did not have an S/N high enough to extract the line-of-sight velocity. The grey-shaded region in the bottom panel represents the area that was not included in the fit to avoid low-S/N areas. The white and grey contours display the location of the dust arc. The beam size is displayed in the bottom left corner. }
    \label{fig:m1_residuals}
\end{figure}

We extracted a 2D line-of-sight velocity map from the clean cube using the package \textsc{bettermoments} and the Gaussian method that fits a Gaussian line profile to each pixel \citep[using a sigma clipping of 5,][]{bettermoments, bettermoments_unc}. The resulting velocity map is presented in the top panel of Figure~\ref{fig:m1_residuals}, displaying a typical rotational pattern.

We fitted a Keplerian model to this velocity map using the \textsc{eddy} package \citep{eddy} to determine the orientation of the system, systemic velocity, and stellar mass. The best-fit values are -3 mas for the RA offset, 20 mas for the Dec offset, 59.6\degree\ for the position angle, 43.8\degree\ for the inclination, 1.89 $M_\odot$ for the stellar mass, and 7.86 km/s for the system line of sight velocity in the Barycentric reference frame. The uncertainties on these values extracted by \textsc{eddy} are below 1\%, but they are likely underestimated as they do not take into account the spatial correlation between pixels. The derived values for the inclination and position angle are consistent with those derived for the dust emission ($44.8\pm0.6\degree$ and $58.5\pm0.8\degree$, respectively). The kinematically derived stellar mass is consistent with the value of 1.901$\pm$0.009 $M_\odot$ presented in \cite{overview_arks}, which was derived based on the available photometry, estimated age \citep[$16\pm2$~Myr,][]{Pecaut2016}, and the PARSEC v1.2S stellar evolution models \citep{bressan2012,Chen2014, Chen2015,Tang2014}. We note that the derived values based on the kinematics could be slightly biased by the potential gas eccentricity of ${\sim}0.03$ with a pericentre towards the SW, as found in Sect.~\ref{sec:profiles}. 

Figure~\ref{fig:m1_residuals} presents the line-of-sight velocity map and residuals after subtracting the best-fit Keplerian model. The dust arc is shown with white and grey contours. We do not find a significant velocity residual feature at the arc location. However, we do find on both ansae (where the projected azimuthal velocity is highest) that the velocities are slower near the outer edge and faster near the inner edge than Keplerian. This effect cannot be explained by an incorrect inclination or stellar mass in the model, since that would not produce residuals with a velocity flip from the inner to the outer regions, which is reversed in the two ansae. Other strong residuals near the inner edge and minor axis (e.g. the strong ${\sim}$200~m/s residual in the NE side) are likely due to the lower S/N near the inner edge and systematic errors due to the finite resolution of the data. They are stronger in lower-resolution cubes (with a robust parameter of 2) and also appear in simulated data (see Appendix \ref{sec:radCO}).

We also explored models with an elevated emission surface as commonly found in protoplanetary discs \citep[e.g.][]{Teague2018}, but we found height values consistent with zero. We note, however, that the emission height for a narrow ring is strongly degenerate with RA and Dec offsets that we fit. Moreover, even if the $^{12}$CO emission is optically thick and elevated, it might not be elevated enough to clearly see in the moment 1 and channel maps, especially if the vertical distribution of the total gas density is narrower than in protoplanetary discs due to a higher mean molecular weight or photodissociation \citep[e.g.][]{Hughes2017, Marino2022, line_arks}. Therefore, for the following analysis, we assume that the emission arises from the midplane.

\subsection{De-projected azimuthal velocity}
\label{sec:kinematics:vrot}

To constrain and deproject the gas velocities as a function of radius, we used the \textsc{eddy} package \citep{eddy}, again using the cube as input. We divided the disc into a set of concentric annuli spaced by 20\% of the beam size (same as the pixel size) and oriented at the same position angle and inclination as the disc (as derived when fitting the line-of-sight velocities). Within each annulus, \textsc{eddy} finds the best azimuthal, radial, and systemic velocity (used to extract the vertical velocity) that reproduces how the centroid velocity changes as a function of azimuth. We used the Gaussian method to derive the line centroid and the `simple harmonic oscillator (SHO)' method to derive the 3D velocities at each radius. The SHO method models the line-of-sight velocity, $v_{\rm los}$, as a function of azimuth within each annuli, as the combined projection of an azimuthal ($v_{\phi}$), radial ($v_{r}$), vertical ($v_{z}$) and systemic velocity ($v_{\rm sys}$),
\begin{equation}
\begin{split}
    v_{\rm los}(r, \phi)=&v_{\phi}(r)\cos(\phi)\sin(i)\\
    - &v_r(r) \sin(r,\phi)\cos(i)\\
    - &v_z(r)\cos(i)+v_{\rm sys}.
\end{split}
\end{equation}
Note that there is a degeneracy between the vertical and systemic velocities. This means that the absolute values of vertical velocities are unconstrained \citep{Teague2019}. Nevertheless, we can fix and approximate the systemic velocity to be the systemic velocity extracted from the Keplerian fit. 

In addition, we also restricted the fit to position angles within 20\degr\ from the major axis of the disc since emission and velocities closer to the minor axis are more affected by the finite beam size due to the disc inclination (44\degree). To account for the uncertainty on the disc orientation, we ran \textsc{eddy} ten times, randomising the disc orientation according to the posterior distribution of the Keplerian fit. For each orientation, \textsc{eddy} performs ten retrievals using a different random sample of pixels within each annulus, which also helps to derive more realistic uncertainties. 

We examined the azimuthal, radial, and vertical velocities, but found that the radial and vertical components are consistent with zero at all radii with upper limits below 1\% of the Keplerian speed. This upper limit applies to global radial or vertical velocities. Locally, radial or vertical velocities could be higher and remain undetected due to the noise. Therefore, hereafter, we focus on the azimuthal component of the velocity.

\begin{figure*}
    \centering
    \includegraphics[width=0.4\linewidth]{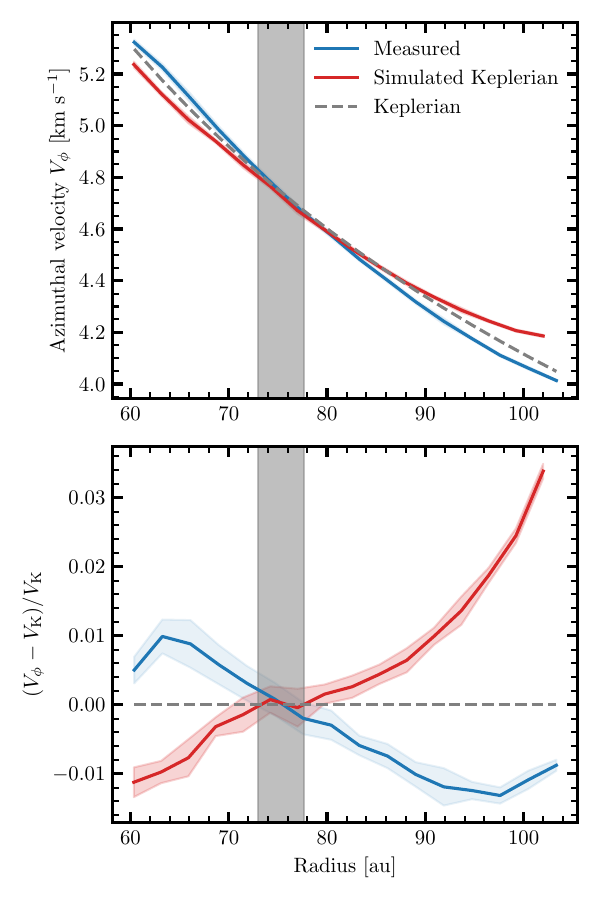}
    \includegraphics[width=0.4\linewidth]{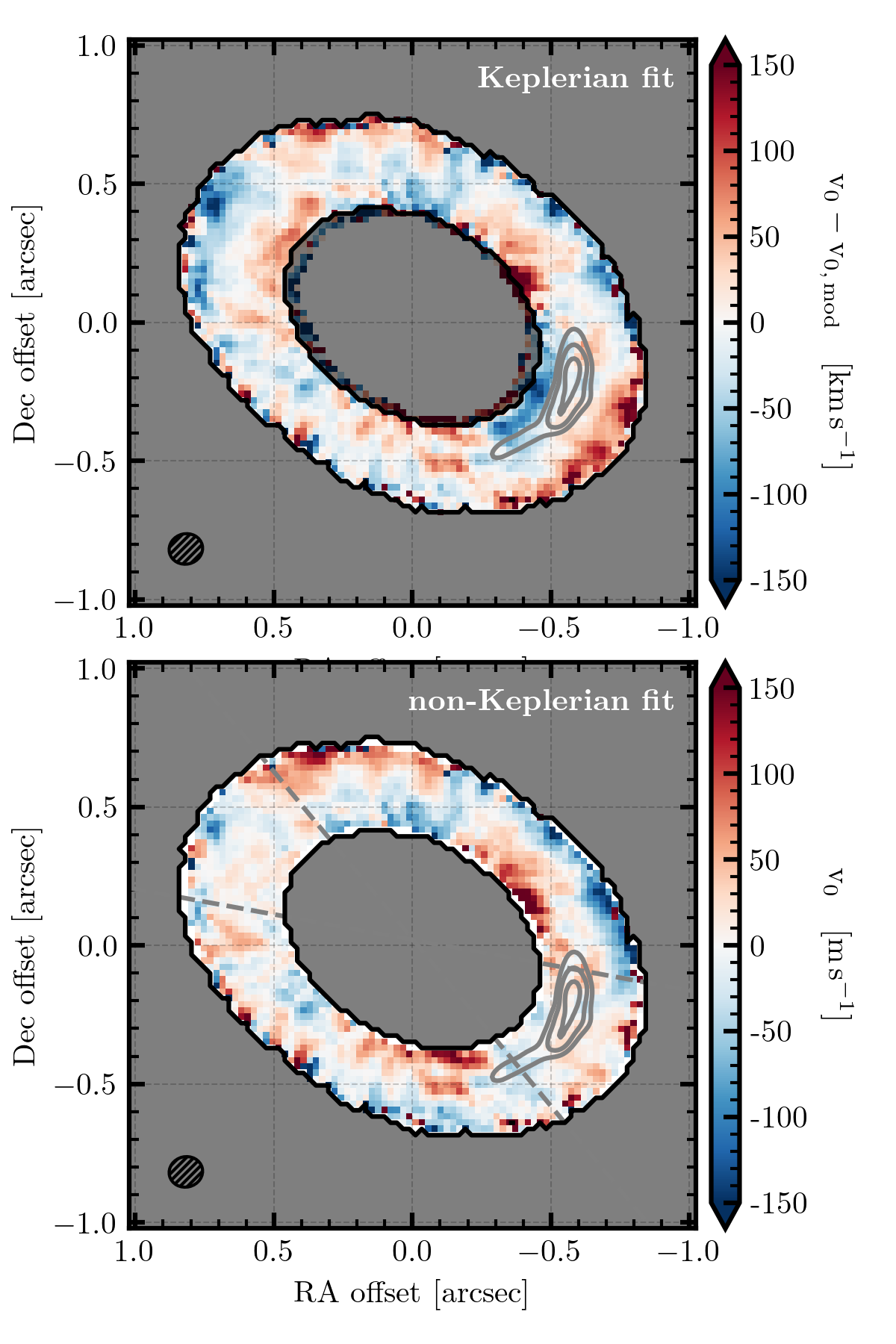}
    \caption{Azimuthal velocity profile fit. Top left: Azimuthal velocity profile extracted from the observed image cube (blue line), from a simulated image cube of a Keplerian model (red line), and the expected Keplerian profile ($v_{\phi}\propto r^{-1/2}$, dashed grey line). Bottom left: Azimuthal velocity profile residual after subtracting the expected Keplerian profile. The blue- and red-shaded regions represent the $1\sigma$ uncertainty. The shaded grey region in the left panels represents the dust arc peak and extent. Top right: Velocity map residuals after subtracting the best-fit Keplerian model. Bottom left: Velocity map residuals after subtracting a model using the measured non-Keplerian azimuthal velocity profile. The dashed grey lines represent wedges used to fit the data near the disc's major axis.  }
    \label{fig:nonkep}    
\end{figure*}

The left top panel in Figure~\ref{fig:nonkep} shows the measured azimuthal velocities (blue) as a function of radius compared to the expected Keplerian velocities (i.e. $v_{\phi}\propto r^{-1/2}$ in grey) and those measured from a simulated observation of a Keplerian model taking into account the effect of the beam and noise (Appendix~\ref{sec:radCO}). The expected Keplerian profile assumes a mass of 1.91~$M_{\odot}$. If the large millimetre-sized grains are radially trapped by the gas due to gas-drag, the dust ring should peak where the gas azimuthal velocity is equal to the Keplerian velocity ($v_{\rm K}$), which occurs for $M_\star=1.91\ M_{\odot}$. The simulated Keplerian model assumes a mass of 1.89~$M_{\odot}$ to roughly match the extracted velocity at 75~au. These values are well in agreement with the $1.90\pm0.01\ M_{\odot}$ inferred using stellar evolution models \citep[as presented in][]{overview_arks}. The bottom left panel shows the residuals after subtracting the expected Keplerian velocities from the observed and simulated data. Overall, we find deviations of the order of 1\% with the azimuthal velocity decreasing with radius faster than Keplerian. This finding is also present if we fit all azimuths, although less strongly, and if we fit the moment 1 instead of the cube. Note that the stellar mass is independently constrained only from stellar models \citep[$1.90\pm0.01\ M_{\odot}$,][]{overview_arks}. Hence, it is uncertain where the extracted azimuthal velocity intersects the Keplerian rotation curve.

The profile extracted from the simulated Keplerian model decreases with radius more slowly than the Keplerian profile. This is due to beam smearing, which contaminates the velocity at the inner and outer edges with the velocities from the peak of emission \citep{Pezzotta2025}. This means that the deviations that we measure from the data are likely even stronger and closer to 2\% the Keplerian speed towards 65 and 85~au.

The right panels in Figure~\ref{fig:nonkep} show the line-of-sight velocity residuals after subtracting the best Keplerian (top) and the more flexible model that fits the 3D velocities as a function of radius (bottom). We find that the latter makes the residuals significantly smaller along the major axis of the disc. The other strong residuals near the disc inner edge and minor axis are still present in the data and are most likely a systematic error due to the limited spatial resolution of the data. For example, the residuals of the simulated Keplerian observations show the same residual pattern near the minor axis (see Figure~\ref{fig:vel_res_model}).

These kinematic deviations are qualitatively what would be expected near a gas ring with a strong pressure maximum at its centre. The azimuthal velocity for a stable and low mass disc ($M_{\rm gas}\ll M_{\star}$) is expected to be
\begin{equation}
    \frac{v_\phi(r,z)^2}{r} = \frac{GM_{\star}r}{(r^2 + z^2)^{3/2}} + \frac{1}{\rho} \frac{\partial P(r,z)}{\partial r}, \label{eq:vrot}
\end{equation}
where $P$ is the gas pressure, which can be defined as 
\begin{eqnarray}
  P &=&  \rho c_s^2,   \\
    c_s^2 &=& \frac{k_{\rm B} T}{\mu m_p},
\end{eqnarray}
where $\rho$ is the gas density, $c_s$ is the sound speed, $T$ is the gas temperature, $\mu$ is the mean molecular weight of the gas, $k_{\rm B}$ is the Boltzmann constant, and $m_p$ is the mass of hydrogen. From Equation~\ref{eq:vrot}, we can see that a positive pressure gradient (e.g. interior to the pressure maximum) results in a super-Keplerian velocity, whereas a negative pressure gradient (e.g. exterior to the pressure maximum) results in a sub-Keplerian velocity. 

The azimuthal velocity profile residuals (bottom left panel in Figure~\ref{fig:nonkep}) also display an apparent dip at 60 and 100~au. These could be due to a change in the slope of the density or temperature profile. However, the low S/N at these radii mean that these features may be due to noise. 

Finally, to test our interpretation of the residual pattern, Appendix~\ref{sec:radCO} presents simulated observations with and without deviations from Keplerian rotation. Our simulations confirm that the observed pattern is consistent with velocity deviations due to the pressure gradient of a gas ring, and inconsistent with a purely Keplerian rotation pattern.  

\subsection{Retrieving the radial density profile}
\label{sec:density_profile}
In this section, we aim to retrieve the gas density profile using the Keplerian deviations as a proxy for the pressure gradient. We start by defining
\begin{equation}
    T = T_0 \bigg(\frac{r}{r_0}\bigg)^\beta, 
\end{equation}
where $T_0$ is a reference temperature at radius $r_0$, and $\beta$ a power law exponent for the temperature. We rearrange Equation~\ref{eq:vrot} with $z=0$ and find
\begin{equation}
    \frac{1}{\rho} \frac{\partial \rho}{\partial r} = \frac{\mu m_p}{k_{\rm B} (\frac{r}{r_0})^\beta} \left(\frac{v_\phi^2}{r T_0} - \frac{GM_{\star}}{r^2 T_0}\right) -\frac{\beta}{r},
\end{equation}
where the quantities on the right-hand side can be estimated from the data ($v_{\phi}$, $T_{0}$, $M_\star$) or assumed to take standard values. We assume that $\beta=-0.14$, $T_{0}=38$~K, and $r_0$=73~au as found in \cite{line_arks} by fitting the $^{13}$CO data cube. Moreover, we assume that $M_\star=1.91\ M_\odot$, which makes the Keplerian speed at the arc location (75~au) the same as the extracted azimuthal velocity. This would be expected if the dust arc corresponds to dust trapped at the pressure maximum. Since we do not know the gas composition, we assume two values for $\mu$: 14 for a disc dominated by C and O (secondary gas), and 2.3 for an H$_2$-dominated disc (primordial gas). Integrating the right-hand side over $r$, we obtain $\ln[\rho(r)]$ plus a constant that becomes a normalisation factor. 

\begin{figure}
    \centering
    \includegraphics[width=1.0\linewidth]{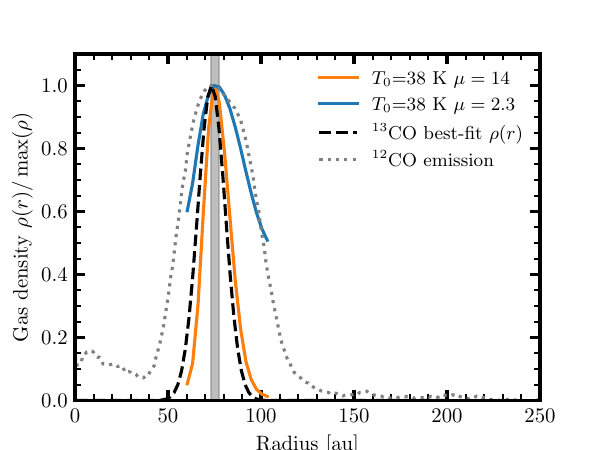}
    \caption{Retrieved gas density profile (solid coloured lines) using different gas temperatures and mean molecular weights. The dashed black line displays the midplane density profile of $^{13}$CO as derived in \cite{line_arks}. The dotted grey line displays the $^{12}$CO gas emission profile (normalised). The vertical shaded grey region shows the location of the arc and dust surface density peak. }
    \label{fig:kin_den}
\end{figure}

Figure~\ref{fig:kin_den} shows the gas density profiles derived from the deviations from Keplerian motion assuming $\mu=14$ (orange) or 2.3 (blue). The dashed black line shows the frequency-integrated CO surface brightness profile for comparison (derived in Sect.~\ref{sec:profiles}). The dashed line shows the best fit midplane density profile of $^{13}$CO derived in \cite{line_arks}. We find a remarkable similarity between the kinematically derived profile assuming $\mu=14$ (orange) and that retrieved from the $^{13}$CO emission (dotted black), with only a slight radial shift that disappears if we assume a stellar mass of 1.92~$M_\odot$. This $\mu$ is also remarkably consistent with the value obtained when fitting the $^{13}$CO cube ($12.6^{+1.3}_{-1.1}$), though see \cite{line_arks} for model assumptions and caveats likely affecting this measurement. The $^{12}$CO emission profile (dotted line) appears much wider than the retrieved density profiles, as expected given its high optical depth.

The kinematically derived profile assuming $\mu=2.3$ (blue) appears ${\sim}3$ times wider than the $^{13}$CO density distribution. This does not, however, rule out the gas having a low mean molecular weight and being primordial. This is because the profile derived from the kinematics traces the total gas distribution, which could differ from that of $^{12}$CO and $^{13}$CO due to CO photodissociation in a primordial or secondary origin scenario \citep{Trapman2019, Marino2020gas}.

Assuming a lower $\mu$ results in a broader density profile because of the sound speed. Decreasing $\mu$ increases the sound speed if we keep the temperature fixed. With a higher sound speed, the density profile can be shallower and still reproduce the same strong pressure gradients and velocity perturbations. Increasing the gas temperature has a similar effect. Decreasing $M_\star$ from 1.91 to 1.89~$M_\odot$ shifts the pressure maximum from 75 to 83~au as $v_{\phi}$ and $v_{\rm K}$ intersect at a larger radius. Increasing $\beta$ shifts the density peak to smaller radii if the density profile is shallow. 

Considering how the beam, optical depth and CO photodissociation affect the CO intensity distribution and azimuthal velocity profiles in different ways, we do not attempt to find a perfect fit between the kinematically derived density and the emission profile. Moreover, an important caveat to note is that the true azimuthal velocity profile is likely steeper than the extracted one because the latter is affected by beam smearing (as is discussed in Sect.~\ref{sec:kinematics:vrot}). This means that the deviations from Keplerian rotation could be underestimated. Nevertheless, our simulated observations presented in Appendix~\ref{sec:radCO} show that our kinematic method can reliably recover the density profile that we input to the model.

In conclusion, we find that the gas kinematics show significant deviations from Keplerian rotation, which can be explained by the pressure gradient of the observed ring of CO gas. Given the degeneracies involved in fitting the density profile, we cannot distinguish between a secondary gas scenario (dominated by C, O or CO) and a primordial one (H$_2$-dominated) using the azimuthal velocity alone. However, if we assume that $^{13}$CO traces well the total gas density distribution, giving consistent results to the $\mu$ measurement in \cite{line_arks}, the CO-dominated scenario would be preferred.

\section{Discussion}
\label{sec:discussion}

\subsection{A vortex as the origin of the arc}
\label{sec:vortex}
We found that the dust overdensity in the millimetre is radially narrow and azimuthally elongated and asymmetric, with the leading side being more azimuthally compact. The dust overdensity is absent or less prominent in the distribution of small grains. All these properties are found in models of vortices in protoplanetary discs \citep[e.g.][]{Baruteau2019}. Gas drag via a vortex can trap dust, creating strong overdensities. This effect is grain size-dependent, which may be the reason why the distribution of small grains does not display a strong overdensity. Grains with frictional timescales comparable to the orbital timescales (i.e. with a Stokes number close to 1) are more efficiently trapped, whereas grains with much shorter frictional timescales tend to follow the gas distribution \citep[e.g.][]{Baruteau2019}. For gas drag to significantly affect the millimetre-sized grains that dominate the observed continuum emission, the gas surface density must be high and comparable to that in protoplanetary discs. If we require millimetre-sized grains to have a fractional timescale of  10 times the orbital period or shorter, i.e. a Stokes number ${\lesssim}10$, this translates to a minimum gas surface density of $5\times10^{-2}$~g~cm$^{-2}$ ($2\times{10^{-3}}\ M_{\oplus}$~au$^{-2}$) or a minimum total gas mass of 20~$M_{\oplus}$ \citep[supported by simulations in][]{vortex_arks}.

This high minimum gas mass of 20~$M_{\oplus}$ may be at odds with the gas surface density if the gas is CO-dominated \citep[estimated to be ${\sim}0.2$~$M_{\oplus}$ depending on the $^{12}$CO/$^{13}$CO abundance ratio,][]{line_arks}. Nevertheless, it is still possible that H$_2$ dominates the gas mass, with an ISM-like H$_2$/CO abundance ratio of $10^{4}$, resulting in a much larger total mass above the minimum required to affect the millimetre-sized dust, and making the gas a protoplanetary disc leftover (i.e. of primordial origin). A vortex in HD~121617 may be a natural consequence of the gas distribution, as vortices can be produced by the Rossby wave instability, which is triggered by strong pressure gradients \citep{Lovelace1999, Li2001, Lyra2009, Meheut2012, Flock2015}. \cite{vortex_arks} explores this scenario using hydrodynamical simulations to constrain what range of gas masses is necessary and to assess whether models can reproduce the amplitude as well as radial and azimuthal extents of the dust arc.

A vortex is also expected to leave a trace in the gas distribution and kinematics. The CO emission distribution does not display any hint of an overdensity. However, there could still be a hidden overdensity in the gas as both $^{12}$CO and $^{13}$CO emissions are optically thick \citep{line_arks}. In addition, as a vortex is a significant local deviation from Keplerian rotation, we may expect to see a feature on the SW side of the disc. The expected local velocity deviations by a vortex may be up to a few \% of the average azimuthal velocity \citep[e.g.][]{Perez2018}, which is at the limit of our sensitivity (the noise in our velocity map residuals is of the order of ${\sim}3\%$ the maximum line of sight velocities). We do not find any significant feature near the dust arc in the residuals after subtracting our non-Keplerian model in Sect.~\ref{sec:kinematics:vrot}. 

Therefore, the observed dust and gas distributions and kinematics are roughly consistent with the presence of a vortex, assuming that there is enough gas to affect the millimetre-sized dust dynamics ($\gtrsim20\ M_{\oplus}$). This would require the gas to be of primordial origin.

\subsection{Narrow gas rings}

The radial distribution of CO gas in debris discs has been challenging to study due to the lack of high spatial resolution observations \citep[e.g.][]{Lieman-Sifry2016, Moor2017}, low S/N \citep[e.g.][]{Marino2016}, and the edge-on orientation of many of these discs, making it harder to extract radial information \citep[e.g. $\beta$~Pic,][]{Matra2017betapic}. ARKS has allowed us to constrain the CO spatial distribution for three non-edge-on gas discs with ages ranging from 16 to 45~Myr  (HD~9672, HD~121617 and HD~131835), and HD~121617 appears to host the narrowest gas disc among these three \citep{gas_arks}. Several of the known gas-rich debris discs may be as narrow, but their edge-on orientation demands a more complex modelling to extract the gas radial distribution \citep[e.g. HD~32297,][]{gas_arks}, or the current observations do not have the resolution to determine this \citep[e.g. HD~9985,][]{Moor2025}. Higher-resolution observations and radiative transfer modelling of those edge-on discs with sufficient observations are needed to constrain how common narrow gas discs are among debris discs. 

Rings of gas (or discs with large cavities) are relatively common among the much brighter protoplanetary discs with ages ranging from 4-20~Myr, and in particular transition discs around early type stars where millimetre-sized grains become radially trapped \citep[e.g.][]{vanderMarel2021, Martinez-Brunner2022, Weber2022}. Moreover, several of those discs display strong asymmetries that resemble the arc in HD~121617 and are interpreted as dust trapping by vortices. If gas in debris discs is primordial, the asymmetry in HD~121617 may be caused by the same processes as in the more massive protoplanetary discs. 

What processes make HD~121617's gas disc narrow is uncertain and may depend on its origin. If gas is primordial, the ring-like distribution of CO gas may be a consequence of photoevaporation carving an increasingly large cavity \citep{Nakatani2021, Nakatani2023, Ooyama2025}, in which case, the different cavity sizes between gas-rich debris discs may be due to different evolutionary stages. Alternatively, the cavity could be produced by multiple planets interior to the disc, which prevents the gas disc from expanding \citep[e.g.][]{Bae2019, Toci2020}. If gas is secondary, the ring-like distribution may be a consequence of a very long viscous timescale or the presence of planets \citep[e.g.][]{Moor2019, Marino2020gas, Kral2020atm}.

\subsection{Vortices in other debris discs}

A vortex has also been proposed to explain an overdensity of CO gas and small dust grains in $\beta$~Pic \citep{Skaf2023}. In that disc, the large dust grains' distribution is more symmetric than the gas \citep[][]{Dent2014, Matra2017, Matra2019} and small grains \citep{Telesco2005, Li2012, Han2023}, which is the opposite of what we found in HD~121617. If the overdensity in both discs is due to vortices, this would suggest that the gas densities in $\beta$~Pic are much lower than in HD~121617 (as expected, given their CO fluxes). A lower gas density would decouple the large millimetre grains from the gas while efficiently trapping the $\mu$m grains in a vortex. Similarly, the HD~181327 debris disc shows an overdensity in the distribution of small grains \citep{Stark2014}, contains sufficient CO gas to affect the dynamics of micron-sized grains (although it is not detected at a sufficient S/N to see any asymmetry), and its millimetre grains are more symmetric \citep{Marino2016}. If this interpretation is correct, then vortices may be common in debris discs with gas and explain dust asymmetries that are grain-size dependent. Determining the gas surface density or kinematics will be crucial to proving or ruling out their presence.

\subsection{A giant collision as the origin of the arc}
\label{sec:coll}

A second possible scenario often invoked to explain overdensities in debris discs is giant collisions \citep[e.g.][]{Wyatt2002, Michel2001, Kenyon2005,  Grigorieva2007, Kral2013, Thebault2018, Jackson2014, Dent2014, Wyatt2016}. Models of (gas-free) giant collisions predict overdensities in the dust distributions that can last for thousands of orbits. Initially, the collision generates a dense cloud of dust that expands and orbits at a roughly Keplerian speed; however, it quickly shears out due to the differing orbital elements of the debris (partly due to radiation pressure acting on small grains), forming a spiral feature. Subsequently, the produced debris returns to the collision point, where new collisions occur, resulting in more debris being generated. The initial clumpy phase is very short, only lasting a few orbits and thus is unlikely to explain the overdensity in HD~121617. 

After a few orbits, the collisional fragments form a smooth but asymmetric disc with a strong overdensity at the collision point that can last for thousands of orbits, or ${\sim0.5}$~Myr for orbits near 75~au \citep{Wyatt2016}. Due to radiation pressure, grains slightly larger than the blow-out size (${\sim}$3--30~$\mu$m) that are produced in the collision point form a disc that is much wider on the opposite side \citep[e.g.][]{Jones2023}. The even smaller grains below the blow-out size that are produced in the collision point are ejected by radiation pressure, forming a leading outward-propagating spiral structure from the collision point \citep{Jackson2014}. The net effect is that scattered light observations tracing small grains should display an asymmetric collision point and a disc that is wider on the opposite side. The distribution of longer-lived millimetre-sized dust traced by ALMA is predicted to be symmetric about the collision point \citep{Jackson2014}.

The key features to assess this scenario are therefore the radial width of the arc, its azimuthal extent and asymmetry, and its apparent absence in the distribution of micron-sized dust and CO gas. The fact that the arc is extremely narrow and potentially unresolved is consistent with the pinch point that occurs at the collision point. The large azimuthal extent could also be consistent with a giant collision scenario. However, the fact that it is asymmetric, with the trailing side being more extended or denser, is inconsistent with the collisional scenario. Moreover, there is no significant overdensity in the distribution of micron-sized dust. Even if the gas affects the dynamics and distribution of the micron-sized dust, larger solids should continuously create small dust at the collision point, generating an asymmetry. Finally, the CO gas does not display any asymmetries, which could be consistent with its continuous release at the collision point. Since the CO gas could be long-lived, its distribution should become axisymmetric as gas pressure would smooth any strong overdensity at the collision point. Additionally, given that the CO gas is very optically thick, an overdensity would be easily hidden.

The interpretation above relies on gas not affecting the dust dynamics. In Sect.~\ref{sec:vortex} we estimated that 20~$M_\oplus$ of gas would be required to significantly affect the dynamics of millimetre-sized grains. A similar calculation leads to a gas mass of 0.02~$M_\oplus$ in gas to affect the dynamics of micron-sized grains. Since the CO gas mass estimated from observations is 0.2~$M_\oplus$ \citep[although there could be more gas][]{line_arks}, the small grains are likely affected by the gas, and thus we cannot use the distribution of small grains alone to rule out this scenario. The CO gas mass could be underestimated, or there could be a large reservoir of H$_2$ gas, in which case even the millimetre-sized grains could be affected by the gas.

The estimated CO gas mass is too large to be consistent with the potential gas released from a giant collision between dwarf or terrestrial planets \citep[e.g.][]{Schneiderman2021}. A collision between gas giant planets may be able to release the required gas, although those collisions are very unlikely as close encounters between gas giants at tens of au are more likely to result in ejections than collisions \citep{Wyatt2017}. Gas could, however, have a different origin from the arc and be dense enough to affect the dust dynamics. Unfortunately, there are no models that predict the evolution of solids after a giant collision in the presence of gas.

In summary, a giant collision scenario by itself is unlikely to explain the multiple observables. However, we cannot rule out the possibility of gas levels being high enough to affect the dynamics of both small and large grains and make the existing giant collision models inapplicable. If that is the case, gas-dust interactions alone may be enough to explain the arc without the need for a giant collision (as discussed in Sect.~\ref{sec:vortex}).

\subsection{Planet-disc interactions as the origin of the arc}

An alternative third scenario to explain the dust asymmetry is the influence of an unseen planet, interacting with the disc through mean-motion resonances (MMRs). Planetesimals and smaller solids trapped in MMRs would form asymmetric structures at specific locations, which could explain the clump in HD~121617. This would require a large number of planetesimals trapped in resonance, which can occur if the planet migrates; this migration causes MMRs to sweep through the disc, trapping large numbers of planetesimals in resonance \citep{Wyatt2003, Wyatt2006, Reche2008, Krivov2007, Booth2023}. This would be the case for a planet located interior to the disc that migrated outwards. This migration could have been driven by planetesimal scattering, akin to Neptune's historical migration into the Kuiper Belt \citep{Malhotra1993}, or by gas interactions in the protoplanetary disc phase. The number of clumps and their shapes depend on the specific MMRs, as well as the planet mass, eccentricity, migration speed and migration distance. This scenario was previously invoked to explain tentative clumps in $\epsilon$ Eri's debris disc \citep{Booth2023}, as well as features in the $\beta$\,Pic and q$^1$\,Eri discs \citep{Matra2019, Lovell2021}. The model will be explored in detail for \mbox{HD 121617} in Pearce et al. (in prep).

Dust trapped in planetary MMRs would have a size-dependent structure. Larger dust would be more strongly constrained in clumps, whilst smaller dust would have a more axisymmetric distribution due to radiation pressure driving grains out of resonance \citep{Wyatt2006}. A similar outcome would arise in the alternative explanation, where the clump is driven by a gas vortex; in the gas scenario, small grains would have a smoother distribution as they are coupled more strongly to the gas \citep{Birnstiel2013}. However, the two scenarios could be observationally distinguished by considering clump rotation. If the clump were resonant, then it would rotate at the orbital speed of the interior planet, which is faster than the Keplerian speed of the disc. Conversely, a vortex clump would rotate at roughly the disc's Keplerian speed \citep{vortex_arks}. Furthermore, a clump generated by a giant impact (Sect.~\ref{sec:coll}) would not rotate at all, because the clump would be caused by the intersection of orbits at the collision point \citep{Jackson2014}. So future observations could distinguish between these scenarios by determining the clump-rotation speed.

\subsection{Non-Keplerian kinematics in debris discs}

This paper presents the first well-resolved non-Keplerian gas kinematics in a debris disc. While these have been frequently found in protoplanetary discs \citep[e.g.][]{Pinte2023}, previous observations of debris discs lacked the sensitivity and resolution to systematically search for non-Keplerian kinematics \citep[e.g.][]{Moor2017}. The only previously known case is the debris disc-bearing class III star, NO Lup, which presents strong evidence of a radial outflow or wind, but remains only marginally resolved by ALMA \citep{Lovell2021}. HD~121617's low inclination and high CO flux make it the ideal ARKS target to study gas kinematics in debris discs, opening the door to future kinematic studies of gas in debris discs over a larger sample of discs with low to moderate inclinations.

In addition to the azimuthal velocity perturbations, gas in debris discs could also present kinematic deviations from Keplerian rotation due to outflows or winds \citep{Lovell2021nolup}, kinks due to planets \citep{Perez2015, Bergez-Casalou2024}, vertical flows due to winds \citep{Nakatani2021}, and vortices \citep{Perez2015}. A future ALMA survey tailored to studying the gas at high spatial and spectral resolutions is needed to assess if these are common features in the gas kinematics of debris discs. Finding any of these features could help us understand the origin and evolution of the gas and the potential presence of planets in these systems.

\subsection{On the origin of HD~121617's gas}

What is the origin of gas in debris discs remains one of the most pressing questions for this field. Depending on its origin, which determines its composition and overall abundance, this gas may affect planets in these systems \citep{Kral2020atm, Marino2020gas}. Typically, secondary models predict gas levels ${\lesssim}10~M_{\oplus}$ that are dominated by CO, carbon, and oxygen, with this upper limit depending on the belt mass, volatile abundances, and gas viscosity \citep{Marino2020gas}. In a primordial scenario, the main constraint is that the CO must be shielded enough by H$_2$ to have a lifetime comparable to the age of the system. This can be achieved by a column density greater than $10^{22}$~cm$^{-2}$ \citep{Visser2009}, which can be translated to a minimum mass of $10\ M_{\oplus}$ if it has a similar radial distribution as CO. This minimum gas mass to shield CO is consistent with the gas mass necessary to trap millimetre-sized grains \citep{vortex_arks}.

In this and the two companion papers by  \cite{line_arks} and \cite{vortex_arks}, we have studied this CO gas-rich system in detail, revealing multiple properties that can help us to assess its origin. Below we summarise the most important:
\begin{itemize}
    \item The $^{12}$CO line is optically thick and thus the total CO mass is unconstrained. Although the $^{13}$CO line is optically thick too, its mass can be constrained to $2\times10^{-3}\ M_{\oplus}$ (in the assumption of LTE and a radially Gaussian, vertically isothermal model), which we can convert to a total CO mass of 0.2~$M_{\oplus}$ by assuming an ISM-like $^{12}$CO/$^{13}$CO abundance ratio of 77 \citep{line_arks}. This CO gas mass could be consistent with both secondary and primordial scenarios, depending on the abundance of shielding species such as CI and H$_{2}$, respectively. \cite{Cataldi2018} constrained the CI mass and column density in this disc, and found it to be insufficient to shield CO, which disfavours a secondary origin. The H$_{2}$ abundance is unconstrained, but if we assume an ISM-like CO/H$_{2}$ abundance ratio of $10^{-4}$, the total H$_{2}$ gas mass would be 200~$M_{\oplus}$, which is sufficient to shield CO for the age of the system.
    \item If the millimetre dust asymmetry is explained by dust trapping, there must be $\gtrsim20\ M_{\oplus}$ of gas to effectively drag the millimetre-sized dust \citep{vortex_arks}. This would favour the gas primordial origin as secondary models struggle to reach those high levels \citep{Marino2020gas}. 
    \item The non-Keplerian kinematics and the density profile fit to the $^{13}$CO emission point towards a high mean molecular weight of ${\sim13}$, favouring a secondary origin. This assumes, however, that the $^{13}$CO traces well the total gas density distribution, which may not be the case due to CO photodissociation.
    \item The peak of the small micron-sized dust distribution is significantly further out than the peak of the millimetre-sized grains, which suggests that gas drag is shaping the distribution of small grains \citep{scat_arks, hd131835_arks}. Both secondary and primordial gas levels could be consistent with this effect \citep{vortex_arks}.
    
\end{itemize}

These features alone and modelled independently cannot be used to reach a definitive conclusion about the origin of the gas for this and other gas-rich debris discs. Additional high spatial and spectral resolution observations of CO and its isotopologues, combined with detailed radiative transfer modelling, are required to reconcile the CO line profiles, moment 0's, line ratios and kinematic deviations. This would give a definitive answer to the optical depths, abundance of CO, kinetic and excitation temperatures, and potentially the mean molecular weight (or sound speed) of the disc. Secondary models of the gas could then be used to reproduce the CO abundance, its radial distribution, and the gas mean molecular weight. Their success or failure could serve as a definitive answer as to the origin of the CO gas.

\section{Conclusions}
\label{sec:conclusions}

In this paper, we have analysed high spatial and spectral resolution ALMA observations of HD~121617 from the large programme ARKS, characterising an asymmetry in the distribution of large millimetre-sized grains and the gas kinematics of this gas-rich exoKuiper belt. In particular, we investigated how strong this asymmetry is in the sub-millimetre continuum emission, whether it is present in the distribution of small grains (traced by scattered light) and gas (traced by CO line emission), and whether the gas displays any non-Keplerian kinematics. Our main results and conclusions are:

\begin{itemize}
    \item At the observed resolution, the arc has a peak intensity at 0.89~mm that is 40\% brighter than the baseline emission around the belt. We do not find any evidence of an arc in the scattered light and CO gas emission. However, we recovered a previous result that the small dust distribution is slightly eccentric \citep{Perrot2023}. We also find evidence that the CO gas and millimetre dust distribution could be eccentric, but this cannot be confirmed with the current astrometric precision.
    \item We fitted a parametric radiative transfer model to the millimetre dust distribution to constrain the arc morphology. The model is composed of an axisymmetric component and an asymmetric arc component. We inferred that the arc component is radially narrow and only marginally resolved with a FWHM between ${\sim}1-5$~au. The azimuthal extent of the arc has a FWHM of ${\sim}90\degree$, with the leading side being more compact than the trailing side. Overall, the arc component is inferred to be 13\% of the total dust mass of $0.2~M_{\oplus}$.
    \item When examining the gas kinematics of the ring, we found that the azimuthal velocity decreases with radius more steeply than Keplerian. This effect can be explained by the density and pressure profile of a narrow ring of gas. Interior to the density and pressure maximum, there is a strong positive pressure that results in super-Keplerian velocities. Exterior to the density and pressure, there is a strong negative pressure gradient that results in sub-Keplerian velocities. We used the extracted azimuthal velocity profile to retrieve the gas density profile, assuming gas molecular weights corresponding to a primordial or secondary origin. Overall, we find that a high molecular weight of 14 (valid for a secondary origin scenario) reproduces best the retrieved $^{13}$CO gas density profile. Assuming a mean molecular weight of 2.3 (expected if the gas is dominated by H$_2$) produces a density profile that is much wider than that of $^{13}$CO. However, our kinematically derived profile traces the bulk of the gas whose distribution could be different to that of CO due to CO photodissociation, skewing its distribution towards the densest parts of the disc. 
    \item We discussed how the arc asymmetry could originate from a vortex in the gas or via planet-disc interactions. In the former case, the arc would have a similar origin to those found in protoplanetary discs that are suspected of being due to dust trapping in a vortex and would require the disc gas mass to be $\gtrsim20~M_{\oplus}$, which would imply a primordial origin for the gas and be enough to shield CO by H$_2$ \citep{vortex_arks}. In the second scenario, we hypothesise that a migrating planet could have trapped solids in mean motion resonances, producing an overdensity in the form of an arc (Pearce et al. in prep).
\end{itemize}

Finally, we note that the HD~121617 arc asymmetry was only found with ARKS' high-resolution and high-S/N observations. Additional arc asymmetries have been found in other ARKS targets \citep{asym_arks} and could be present in many other exoKuiper belts not included in ARKS.
    
\section*{Data availability}

The ARKS data used in this paper and others can be found in the \href{https://dataverse.harvard.edu/dataverse/arkslp}{ARKS dataverse}. The continuum image used in this work can be found within the \href{https://dataverse.harvard.edu/dataset.xhtml?persistentId=doi:10.7910/DVN/VNGHPQ&}{ARKS I's dataset} (\href{https://doi.org/10.7910/DVN/VNGHPQ}{doi.org/10.7910/DVN/VNGHPQ}). The CO gas images used in this work can be found within the \href{https://dataverse.harvard.edu/dataset.xhtml?persistentId=doi:10.7910/DVN/PXGNNZ&}{ARKS IV's dataset} (\href{https://doi.org/10.7910/DVN/PXGNNZ}{doi.org/10.7910/DVN/PXGNNZ}). The SPHERE scattered light image used in this work can be found within the \href{https://dataverse.harvard.edu/dataset.xhtml?persistentId=doi:10.7910/DVN/RGRKTJ&}{ARKS V's dataset} (\href{https://doi.org/10.7910/DVN/RGRKTJ}{doi.org/10.7910/DVN/RGRKTJ}). For more information, visit \href{https://arkslp.org}{arkslp.org}.

\bibliographystyle{aa}
\bibliography{bib}

\appendix

\section{Acknowledgments}

\begin{acknowledgements}

We thank the anonymous referee for their comments and suggestions that have improved the quality and clarity of this paper. We thank Richard Teague for helpful discussions and for providing help using the packages \textsc{bettermoments}, \textsc{eddy} and \textsc{gofish}. 
SM acknowledges funding by the Royal Society through a Royal Society University Research Fellowship (URF-R1-221669) and the European Union through the FEED ERC project (grant number 101162711). PW acknowledges support from FONDECYT grant 3220399 and ANID -- Millennium Science Initiative Program -- Center Code NCN2024\_001. TDP is supported by a UKRI Stephen Hawking Fellowship and a Warwick Prize Fellowship, the latter made possible by a generous philanthropic donation. AB acknowledges research support by the Irish Research Council under grant GOIPG/2022/1895. SP acknowledges support from FONDECYT Regular 1231663 and ANID -- Millennium Science Initiative Program -- Center Code NCN2024\_001. SMM acknowledges funding by the European Union through the E-BEANS ERC project (grant number 100117693), and by the Irish research Council (IRC) under grant number IRCLA- 2022-3788. Views and opinions expressed are however those of the author(s) only and do not necessarily reflect those of the European Union or the European Research Council Executive Agency. Neither the European Union nor the granting authority can be held responsible for them. LM acknowledges funding by the European Union through the E-BEANS ERC project (grant number 100117693), and by the Irish research Council (IRC) under grant number IRCLA- 2022-3788. Views and opinions expressed are however those of the author(s) only and do not necessarily reflect those of the European Union or the European Research Council Executive Agency. Neither the European Union nor the granting authority can be held responsible for them. JM acknowledges funding from the Agence Nationale de la Recherche through the DDISK project (grant No. ANR-21-CE31-0015) and from the PNP (French National Planetology Program) through the EPOPEE project. CdB acknowledges support from the Spanish Ministerio de Ciencia, Innovaci\'on y Universidades (MICIU) and the European Regional Development Fund (ERDF) under reference PID2023-153342NB-I00/10.13039/501100011033, from the Beatriz Galindo Senior Fellowship BG22/00166 funded by the MICIU, and the support from the Universidad de La Laguna (ULL) and the Consejer\'ia de Econom\'ia, Conocimiento y Empleo of the Gobierno de Canarias. EC acknowledges support from NASA STScI grant HST-AR-16608.001-A and the Simons Foundation. AMH acknowledges support from the National Science Foundation under Grant No. AST-2307920. MRJ acknowledges support from the European Union's Horizon Europe Programme under the Marie Sklodowska-Curie grant agreement no. 101064124 and funding provided by the Institute of Physics Belgrade, through the grant by the Ministry of Science, Technological Development, and Innovations of the Republic of Serbia. This work was also supported by the NKFIH NKKP grant ADVANCED 149943 and the NKFIH excellence grant TKP2021-NKTA-64. Project no.149943 has been implemented with the support provided by the Ministry of Culture and Innovation of Hungary from the National Research, Development and Innovation Fund, financed under the NKKP ADVANCED funding scheme. JBL acknowledges the Smithsonian Institute for funding via a Submillimeter Array (SMA) Fellowship, and the North American ALMA Science Center (NAASC) for funding via an ALMA Ambassadorship. EM acknowledges support from the NASA CT Space Grant. A.A.S. is supported by the Heising-Simons Foundation through a 51 Pegasi b Fellowship. Support for BZ was provided by The Brinson Foundation.
This paper makes use of the following ALMA data: ADS/JAO.ALMA\# 2022.1.00338.L. ALMA is a partnership of ESO (representing its member states), NSF (USA) and NINS (Japan), together with NRC (Canada), MOST and ASIAA (Taiwan), and KASI (Republic of Korea), in cooperation with the Republic of Chile. The Joint ALMA Observatory is operated by ESO, AUI/NRAO and NAOJ. The National Radio Astronomy Observatory is a facility of the National Science Foundation operated under cooperative agreement by Associated Universities, Inc. The project leading to this publication has received support from ORP, which is funded by the European Union’s Horizon 2020 research and innovation programme under grant agreement No 101004719 [ORP]. We are grateful for the help of the UK node of the European ARC in answering our questions and producing calibrated measurement sets. This research used the Canadian Advanced Network For Astronomy Research (CANFAR) operated in partnership by the Canadian Astronomy Data Centre and The Digital Research Alliance of Canada with support from the National Research Council of Canada the Canadian Space Agency, CANARIE and the Canadian Foundation for Innovation.
\end{acknowledgements}

\section{Posterior distribution of the dust arc model}

\begin{figure*}[h]
\centering
\includegraphics[width=0.99\textwidth]{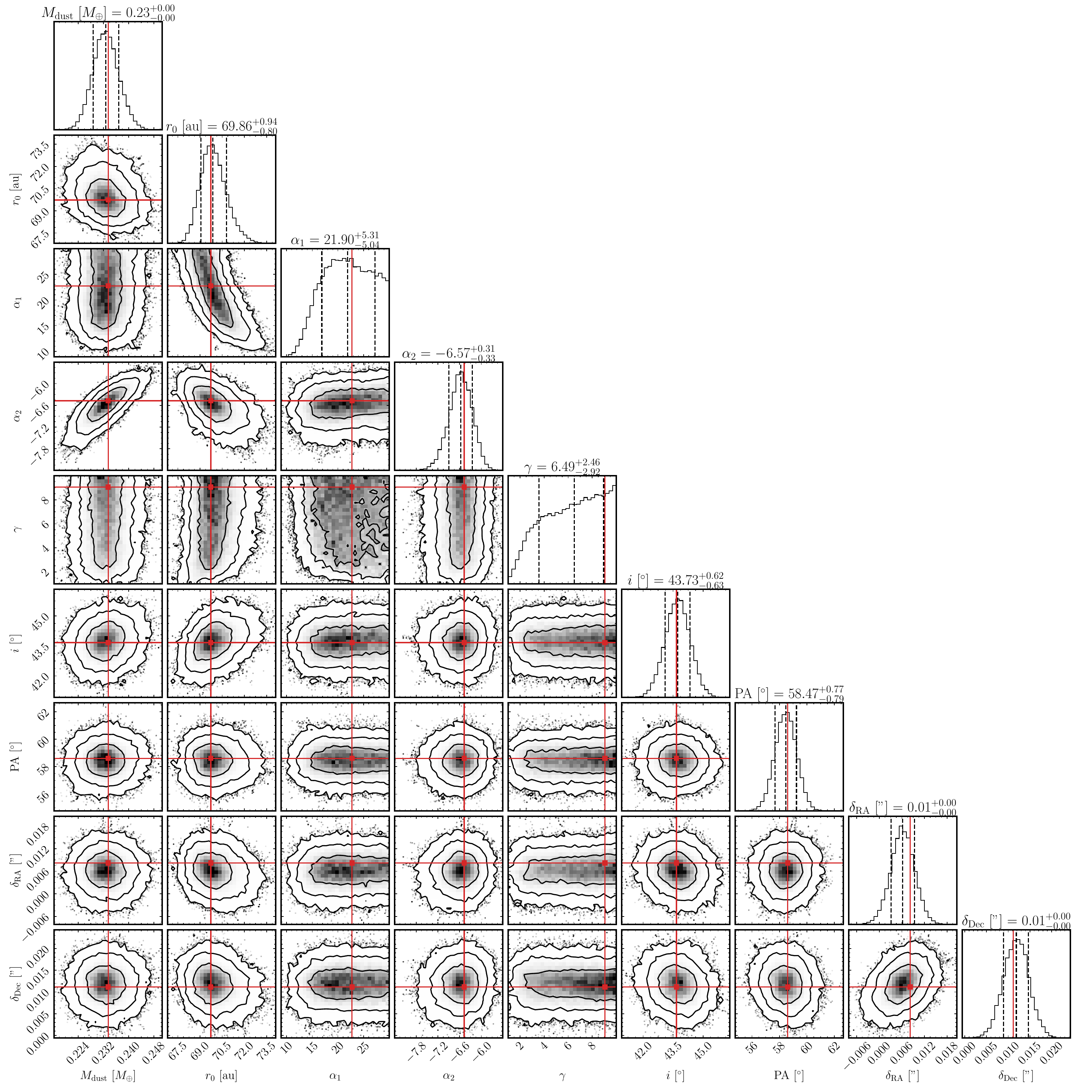}
\caption{MCMC posterior distribution of the parameters used to fit the dust continuum observations, except those that define the arc. The contour levels in the 2D marginalised distributions correspond to the 68, 95, and 99.7\% confidence levels. The dashed vertical lines in the marginalised distributions display the 16th, 50th, and 84th percentiles. The red lines represent the best-fit value (lowest $\chi^{2}$).}
    \label{fig:full_mcmc}
\end{figure*}

Figure~\ref{fig:full_mcmc} shows the posterior distribution retrieved from the MCMC fit to the dust ALMA data, except those that define the arc, which are shown in Figure~\ref{fig:mcmc}.

\section{$^{12}$CO radiative transfer models}
\label{sec:radCO}

To understand the systematics in the velocity profiles derived in Sect.~\ref{sec:kinematics}, we simulate $^{12}$CO J=3-2 ALMA observations using Keplerian and non-Keplerian models to test the sensitivity of our methods. We start by simulating synthetic image cubes using \textsc{disc2radmc} \citep{Marino2022}. Our model consists of a disc of gas around a 1.91~$M_{\star}$ central star. The gas is distributed with a Gaussian radial surface density distribution, centred at 74~au, with a FWHM of 17~au \citep{line_arks}, and a vertical aspect ratio $h=0.05$. 

The gas is assumed to be in local thermodynamic equilibrium, and we set the kinetic and excitation temperature at the belt centre to be 38~K and decaying with radius with an exponent $\beta$ of -0.14 \citep[as derived in ][]{line_arks}. 
The mean molecular weight ($\mu$) is assumed to be either 14 (representative of a secondary origin scenario) or 2.3 (primordial scenario). A lower $\mu$ results in a higher sound speed and larger deviations from Keplerian rotation.
The model azimuthal velocities are set to be Keplerian or non-Keplerian rotation due to pressure support (following Eq. \ref{eq:vrot}). The radial and vertical velocities are set to zero.

We simulate image cubes with a spectral resolution of 26~m/s and spatial resolution of 10~mas. We use these synthetic cubes to simulate ALMA observations using the task \textsc{simobserve} in CASA \citep{casa}, using the same antenna configuration as the observations and inputting the same noise level per channel. The simulated ALMA observations are then imaged using \textsc{tclean} using a Briggs weighting, a robust parameter of 0.5, a Keplerian mask and the same pixel size as in \cite{gas_arks}. The resulting image cubes have approximately the same beam size and root-mean-square level.
Finally, we follow the same procedure as with the real data, and extract the line-of-sight velocities, residuals after a Keplerian model subtraction, and azimuthal velocities from the simulated observations. 

\begin{figure*}
\centering
    \includegraphics[trim=0.0cm 0.2cm 1.5cm 0.2cm,
    clip=true, width=0.27\textwidth, valign=tl]{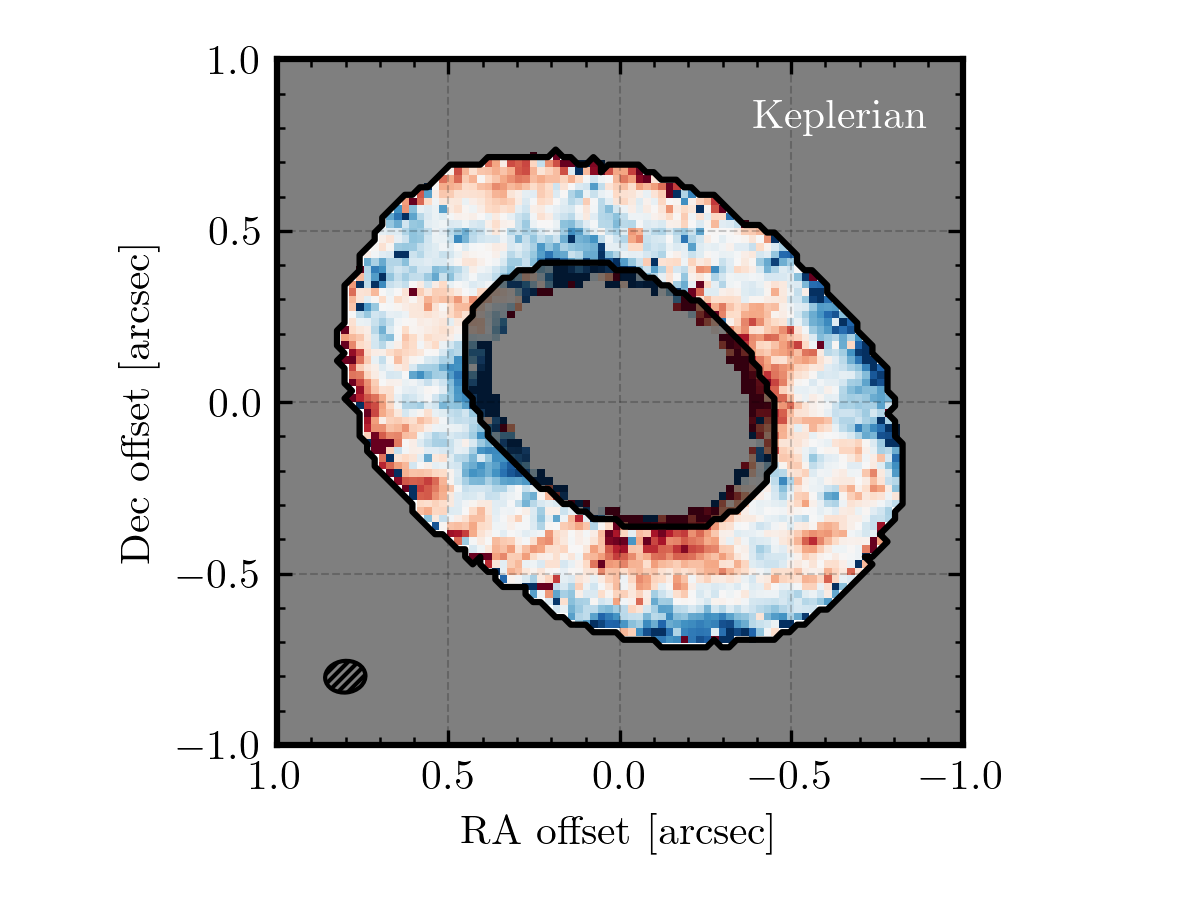}
    \includegraphics[trim=1.5cm 0.2cm 1.5cm 0.2cm,
    clip=true,width=0.222\textwidth, valign=tl]{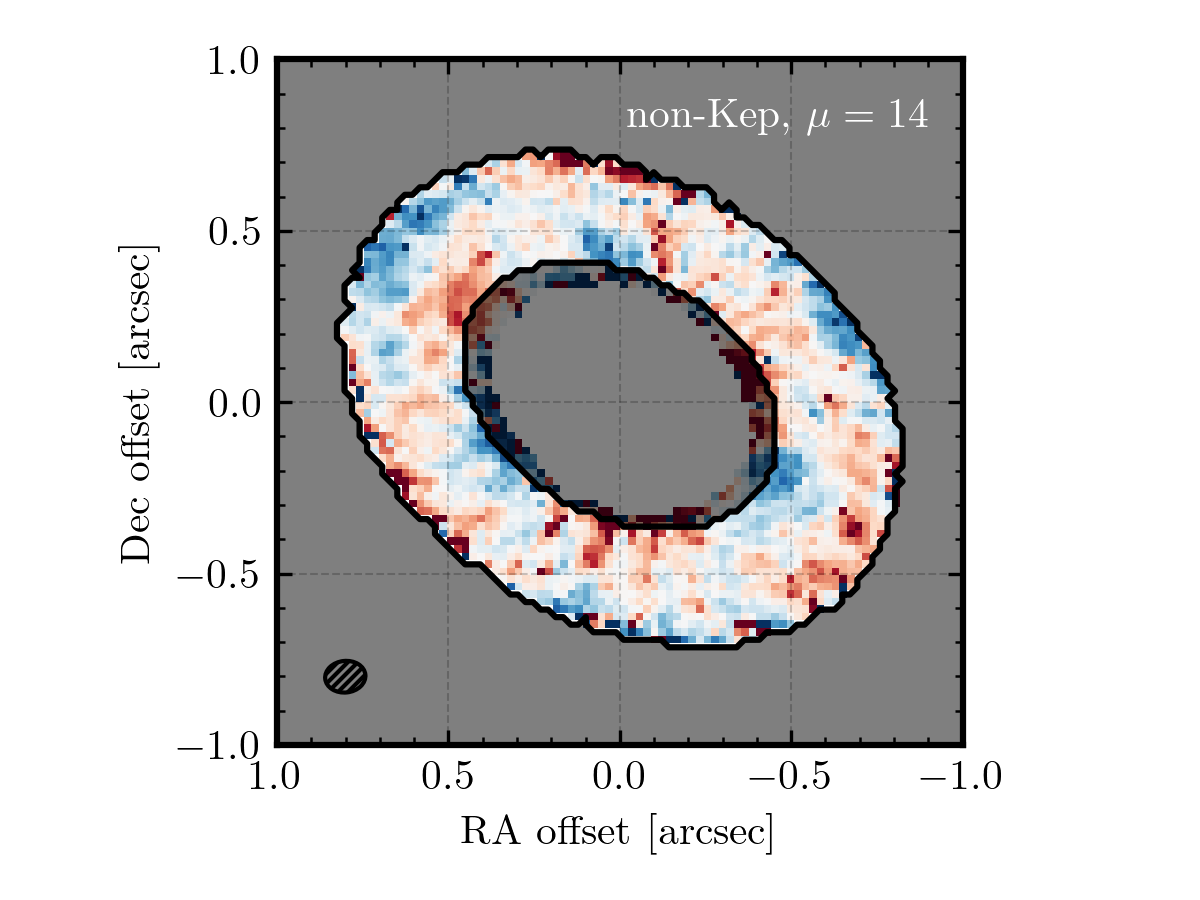}
    \includegraphics[trim=1.5cm 0.2cm 1.5cm 0.2cm,
    clip=true,width=0.222\textwidth, valign=tl]{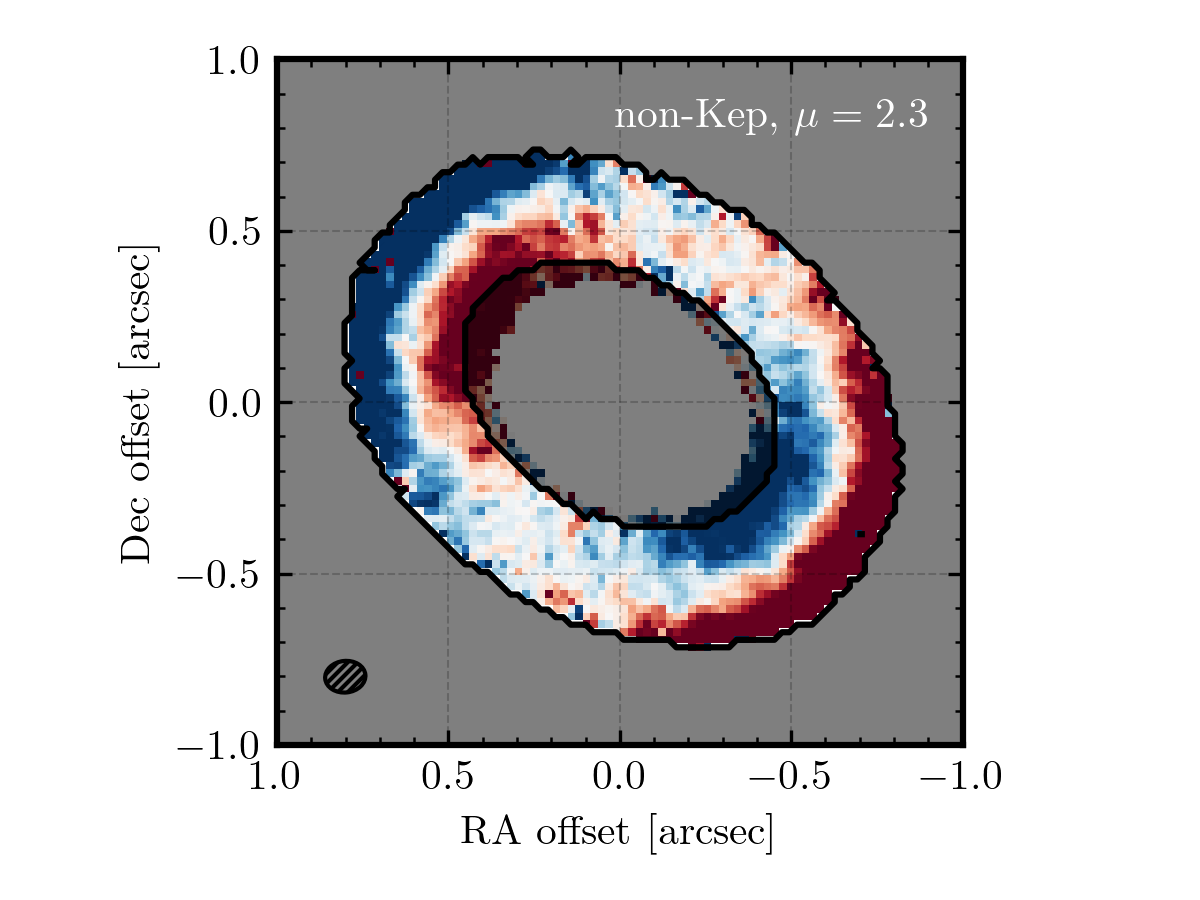}
        \includegraphics[trim=1.5cm 0.2cm 0.2cm 0.2cm,
    clip=true,width=0.26\textwidth, valign=tl]{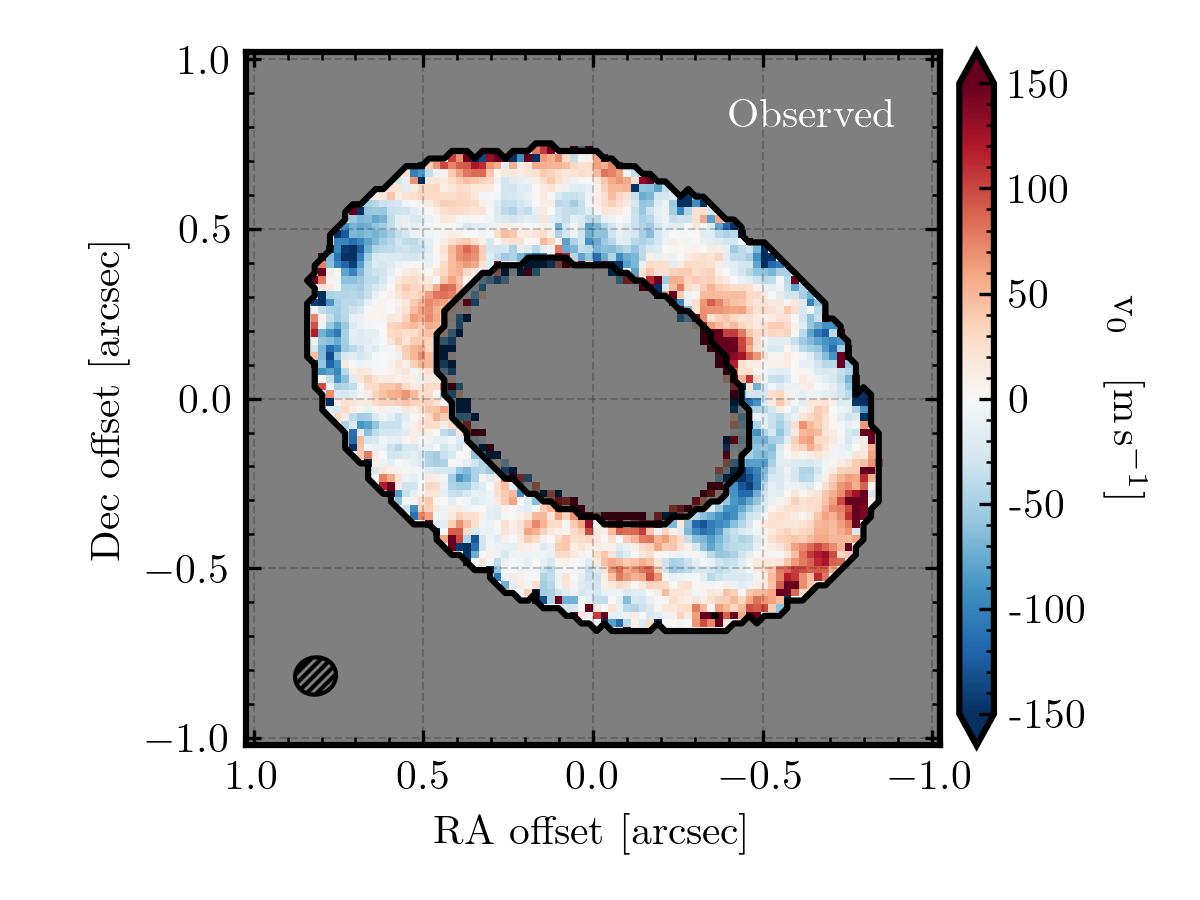}
    \caption{Velocity map residuals after subtracting the best fit line-of-sight Keplerian model from simulated (left, middle left and middle right panels) and the real observations (right panel). The grey-shaded region represents the area that was not included in the fit to avoid low S/N areas. The beam size is displayed in the bottom left corner. } \label{fig:vel_res_model}
\end{figure*}

The line-of-sight velocity residuals for three different models and the observations are presented in Figure~\ref{fig:vel_res_model}. The left panel presents a Keplerian model, which does not display any clear residual pattern along the disc major axis as in our observations. The central panels present two non-Keplerian models. The middle left and right panels present non-Keplerian models of a disc with a high $\mu$ of 14 and one with a low value of 2.3, respectively. Both simulated models display a similar pattern to the observations, with the inner NE (SW) region being more red-shifted (blue-shifted) than the Keplerian model, indicating a super-Keplerian velocity. Similarly, the outer regions display sub-Keplerian velocities. The residual velocities of the observations are similar to those in the high $\mu$ model, but much lower than those in the low $\mu$ model. 

\begin{figure*}[h!]
\centering
    \includegraphics[trim=0.0cm 0.0cm 0.0cm 0.0cm,
    clip=true, width=1.0\textwidth]{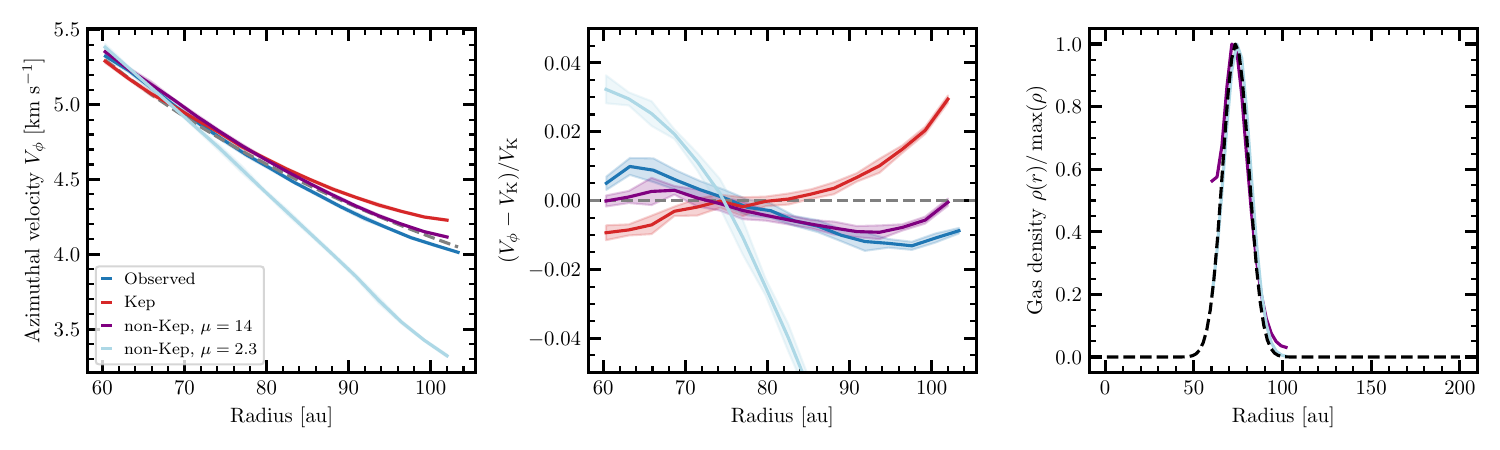}
    \caption{Left panel: Azimuthal velocity profile extracted from the simulated and real image cubes vs the best-fit Keplerian model of the real data (dashed grey line). Middle panel: Azimuthal velocity profile residual after subtracting a Keplerian profile. The shaded regions represent the 1$\sigma$ uncertainty. Right panel: Gas density extracted from the two models with non-Keplerian velocities (solid coloured lines) compared to the input gas density (dashed black line). The dark blue line represents the profiles extracted from the real data. The light blue, purple, and red lines represent different models as indicated in the legend in the top panel. } \label{fig:vrot_models}
\end{figure*}

To better compare the simulated models and data, the top panel of Figure~\ref{fig:vrot_models} shows the azimuthal velocities extracted using the same method as for the data. The velocities extracted from the simulated observation of a Keplerian model (red line) decrease with radius more slowly than both the observations (dark blue line) and a Keplerian profile (dashed grey line). The latter is likely due to beam smearing \citep{Pezzotta2025}. The two profiles extracted from non-Keplerian simulated models (purple and light blue) have slopes that are steeper than Keplerian, as in the observations. The best-fit $M_{\star}$ for the Keplerian, and non-Keplerian $\mu=14$, and $\mu=2.3$ models was 1.93, 1.93 and 1.80 $M_{\odot}$, significantly different from the 1.91~$M_{\odot}$ used as input. This demonstrates that the stellar mass derived from Keplerian fits can be easily wrong by a few per cent.

The middle panel of Figure~\ref{fig:vrot_models} displays the azimuthal velocity residuals of the simulated and real observations after subtracting a Keplerian profile that intersects the corresponding azimuthal velocities at 75~au. The two non-Keplerian simulated observations (purple and light blue lines) produce residuals with the same pattern as the observations (dark blue), demonstrating how the effect of pressure support can lead to the non-Keplerian signal that we found. The deviations from the model with a low $\mu$, and thus a high sound speed, are much stronger than observed. We note, however, that there are strong degeneracies between the density distribution of the gas and its sound speed (determined by its temperature and $\mu$); therefore, we do not attempt to find a best fit here.

The Keplerian simulated observations produce a deviation opposite to what is observed, with sub-Keplerian at the inner edge and super-Keplerian at the outer edge due to beam smearing. This demonstrates that a Keplerian model is strongly disfavoured by our observations. 

The bottom panels show the inferred gas density radial profiles extracted from the two models with non-Keplerian velocities. We use as input the known model gas temperature, mean molecular weight, and a stellar mass such that the retrieved velocities match the Keplerian velocity at 75~au. For both non-Keplerian models, the derived profiles reproduce well the true input profile despite the effects of beam smearing and optical depth effects. 

However, if we use as input the stellar mass of the radiative transfer model of $1.91~M_{\odot}$, the retrieved profiles do not recover as well the input density. For $\mu=2.3$, the peak density shifts to 70~au. For $\mu=14$, the retrieved profile monotonically rises with radius, creating a positive pressure gradient that is necessary to match the observed azimuthal velocity that is higher than Keplerian at all radii. The difference in behaviour between these two models is likely due to the velocity deviations being stronger and easier to extract for the $\mu=2.3$ model.

The examples shown in this section demonstrate that the non-Keplerian kinematics of HD~121617 can be reasonably reproduced by radiative transfer models that incorporate the effect of pressure support on the azimuthal velocity of the gas. While we do not attempt to fit the data with these models due to strong degeneracies between model parameters, this could be done, in principle, in the future by incorporating other observables such as the line profile (constraining the gas temperature) and intensity distribution (constraining the spatial distribution).

\end{document}